\documentclass[smallextended]{svjour3}       
\smartqed  
\usepackage{graphicx}

\usepackage{amssymb}
\usepackage{longtable}
\usepackage{supertabular}
\usepackage{multirow}
\usepackage[version=3]{mhchem} 
\usepackage[usenames, dvipsnames]{color}
\usepackage{xcolor}
\usepackage{array}
\usepackage[utf8]{inputenc}
\usepackage{subfigure}
\usepackage{float}

\begin{document}
\title{Electronic excited states of benzene in interaction with water clusters : influence of structure and size}
\subtitle{Time dependent density functional theory {\it vs} multireference wavefunction approaches.}

\author{ Nadia Ben Amor       \and
         Eric Michoulier  \and Aude Simon
}

\institute{Nadia Ben Amor \at
             Laboratoire de Chimie et Physique Quantiques (LCPQ), F\'ed\'eration FERMI, Univ. Toulouse UT3 \& CNRS, UMR5626, 118 Route Narbonne, F-31062 Toulouse, France \\
           \and
           Eric Michoulier \at
               Laboratoire de Chimie et Physique Quantiques (LCPQ), F\'ed\'eration FERMI, Univ. Toulouse UT3 \& CNRS, UMR5626, 118 Route Narbonne, F-31062 Toulouse, France \\
        \and Aude Simon \at
        Laboratoire de Chimie et Physique Quantiques (LCPQ), F\'ed\'eration FERMI, Univ. Toulouse UT3 \& CNRS, UMR5626, 118 Route Narbonne, F-31062 Toulouse, France \\
        \email{aude.simon@irsamc.ups-tlse.fr} 
}

\date{Received: date / Accepted: date}

\maketitle
\begin{abstract}
    This work is dedicated to the theoretical investigation of the influence of water clusters' organisation and size on the electronic spectrum of an interacting benzene (Bz) molecule using both TD-DFT and CASPT2 approaches. Two series of geometries, namely $Geo_{IEI}$ and $Geo_{IED}$ were extracted from two Bz-hexagonal ice configurations leading to maximum and minimum ionization energies respectively. An appropriate basis set containing  atomic diffuse and polarisation orbitals and describing the Rydberg states of Bz was determined. The TD-DFT approach was carefully benchmarked against CASPT2 results for the smallest systems.
    Despite some discrepancies, the trends were found to be similar at both levels of theory: the positions and intensities of the main $\pi \rightarrow \pi^{\star}$ transitions were found slightly split due to symmetry breaking.
    For the smallest systems, our results clearly show the dependence of the electronic transitions on the clusters' structures. Of particular interest, low energy transitions of non negligible oscillator strength from a Bz $\pi$ orbital to a virtual orbital of Rydberg character, also involving atomic diffuse functions and partially expanded on the water cluster, were found for the $Geo_{IED}$ series. The energies of such transitions were determined to be more than 2\,eV below the ionization potential of Bz.  When the cluster's size increases, similar transitions were found for all structures, the virtual orbitals becoming mainly developed on the H atoms of the water molecules at the edge of the cluster. Given their nature and energy, such transitions could play a role in the photochemistry of aromatic species in interaction with water clusters or ice, such processes being of astrophysical interest.
     \\
     
\end{abstract}

\section{Introduction}

The study proposed in this paper is part of an ongoing effort aiming at understanding the physical and chemical processes following the photoactivation of  carbon matter in interstellar space. In space, carbon is the fourth most abundant element and it is mainly present under the form of large carbonaceous molecules \cite{dartois_interstellar_2019}. Among them, those possessing an aromatic character, polycyclic aromatic hydrocarbons (PAHs),  have received considerable interest  since they were proposed, in the mid-eighties,  as the carriers of  the aromatic infrared bands (AIBs), a set of mid-IR emission bands observed in many regions of the interstellar medium \cite{aib2,leger84}. This 'Astro-PAH' population would constitute  about 10 to 20\,\% of the total elemental carbon in the interstellar medium (ISM) \cite{draine2003,joblin1992} although no specific PAH molecule has been identified yet \cite{PAHbook}. Astro-PAHs may also play an important role in the chemistry of the  ISM,  sustaining interest for the investigation of their energetic processing  (see \cite{sprchapter_2017} and references therein).  \\

In dense interstellar clouds, where temperatures are low ($<$ 50~K), molecular PAHs may condense onto the icy mantles of dust grains \cite{Bowman2011}.  UV irradiation or energetic particle bombardment of these dust particles are expected to induce energetic processes leading to a specific chemistry within the ice \cite{chemRev2016}. This astrophysical context motivated experimental studies investigating the photoprocessing of PAHs in water ices upon UV irradiation under different conditions. Under high energy VUV irradiation, the formation of PAH cations was observed \cite{Bernstein2007,Bowman2010,Bouwman2009}, as well as a rich  photochemistry, leading to the formation of alcohols, quinones and ethers \cite{Bernstein1999,Bernstein2007,Bernstein2001}. At lower energy ($\lambda$ $>$ 235\,nm), the photoreactivity of pyrene and coronene with water molecules was  observed in argon matrix or in amorphous solid water (ASW) at 10~K \cite{Guennoun2011,Guennoun2011bis,barros17} also revealing the production of oxygen-containing PAH products. 
Interestingly, this reactivity occurs at energies below the ionisation energy (IE) of isolated PAHs ($\sim$7.4 and 7.2\,eV for pyrene and coronene, respectively). 
One hypothesis is that photoreactivity, even at lower energy, could be ion-mediated \cite{Bouwman2009}. 
Experimental studies, complemented with modelling, showed that the ionisation of PAHs adsorbed on water ice required about 1.5 to 2.0\,eV less energy than in the case of isolated PAHs \cite{Gudipati2004,Woon_2004}. We showed using explicit description of the electrons within the  self consistent charge density functional based tight binding (SCC-DFTB) scheme \cite{dftb1,dftb2,elstner98}, that this could not be accounted for by the lowering of the IE, which was found less than 1\,eV \cite{michoulier_theoretical_2018}. 
A way to rationalize the experimental results would be to take into account the fact that the electrons released by the ionisation could attach to $\mathrm{H_2O}$ molecules or recombine with free radicals such as OH formed during ice photolysis \cite{Gudipati2006} and to subtract the electron affinity of the electron receptor. An alternative hypothesis is the formation of a charge transfer (CT)  PAH$^+$-(H$_2$O)$_n^-$ 
electronic excited state \cite{Noble2017} of sufficiently long lifetime to be detected and react.  \\

In order to get insights into the possibility of formation of such state, we investigated the electronic excited states of a benzene molecule solvated by a series of (H$_2$O)$_n$  clusters (n $<$ 50), benzene being the simplest aromatic carbon molecule (quoted C$_6$H$_6$ or Bz hereafter). 
The vertical electronic spectra were computed at the time dependent density functional theory (TD-DFT) and multi-state complete active space with perturbation theory at the second order (MS-CASPT2) \cite{finley_multi-state_1998} levels for the smallest clusters. The comparison between both approaches is mandatory, the  presence of Rydberg and charge-transfer states in these systems requiring the use of adequate functionals and basis sets. \\

A few theoretical studies were dedicated to the investigation of the effect of the interaction of water clusters with Bz on its electronic spectrum. Vertical TD-DFT spectra of Bz(H$_2$O)$_6$ were computed with two conformations of the water hexamer, cage and prism \cite{sharma_structure_2014}, showing a red-shift of the $\pi \rightarrow$ $\pi^{*}$ transition of Bz. From their study, they conclude that the benzene interaction with (H$_2$O)$_6$
cluster plays a significant role in giving new excitation features
in the UV spectra of Bz–(H$_2$O)$_6$ clusters, and that charge transfer (CT) states and locally excited diffuse states play an important role in such systems. Interestingly, they show that the obtained electronic spectrum is dependent on the functional, and that MO6-2X and CAM-B3LYP generally show good agreement. They used the 6-31++G(d,p) basis set that is not expected to describe properly the Rydberg states of Bz but the authors do not focus on the correct description of the latter states.  Sharma {\it et al.} \cite{Sharma2016_bzIce} pursued their study expanding the benzene environnment. They computed TD-DFT spectra of Bz adsorbed on hexagonal (Ih) water ice using the ONIOM(QM:MM) formalism with the QM part at the MO62X/6-31++G(d,p) level and the MM part using AMOEBA09. Their conclusions regarding the electronic spectra was quite similar to their previous work \cite{sharma_structure_2014}, also showing the presence of Bz-mediated transition in the spectrum of water ice. Besides, they highlighted the dependence of the IE values and electronic spectra on the interface structure {\it ie} the presence of dangling OH (free OH on the ice surface). A few years later, Michoulier {\it et al.} \cite{michoulier_adsorption_2018} generalized this study showing the dependence of the PAH-ice interaction energy on the structure at the interface for a series of PAHs and ice types using a molecular dynamics/force field approach. 
This dependence was also shown through the SCC-DFTB scheme \cite{michoulier_theoretical_2018}.  \\

The influence of the environment constituted of water clusters on the electronic spectrum of carbonaceous aromatic species was also studied for small PAHs using quantum chemical calculations.  Sharma {\it et al.} published a similar study as that for Bz–(H$_2$O)$_6$ \cite{sharma_structure_2014} replacing benzene by naphthalene \cite{sharma_ground_2015}. The  influence of a water ice environment (modelled by  about 40  water molecules)  on the vibronic profile of the absorption spectrum and on the fluorescence spectrum  of pyrene below 4.5\,eV was  studied through {\it ab initio}  simulations \cite{pyr_wat2014}. \\

In the present study, we report TD-DFT and MS-CASPT2 results for two series of Bz-(H$_2$O)$_n$  clusters (n $<$ 50), differing by the organisation of the water molecules which, for one series, lead to an increase of the IE, and for the other one to a decrease of the IE. The geometries of the clusters, obtained at the SCC-DFTB level, were extracted from those determined in previous work \cite{michoulier_theoretical_2018} (see subsection \ref{subsec:comp1}). 
We showed in that study that 
an increase of the IE was due to the presence of interacting dangling OH while a decrease of the IE was due to the presence of interacting water oxygens.  In the present work, we aim at determining the influence of the structural variations on the electronic spectrum, searching for a correlation between a decrease of the IE and the presence of low lying excited states involving charge transfer from Bz to the interacting water cluster. \\

This article is organized as follows : the computational details including the benchmark on isolated Bz are reported in Section \ref{sec:comp}. The results for the two series of Bz-water clusters increasing the size of the water clusters are reported and discussed in  Sections \ref{sec:res} and \ref{sec:disc}. Finally some conclusions are drawn in Section \ref{sec:conc}. \\

\section{Computational details}\label{sec:comp}
\subsection{Systems}\label{subsec:comp1}

\begin{figure}
        \centering
        \includegraphics[width=10cm]{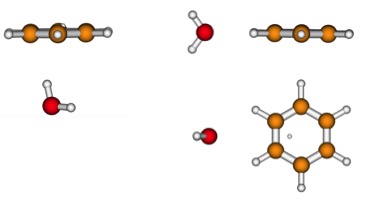} \\
        (a) \hspace{4.5cm} (b) \\
        \caption{DFTB locally optimized geometries of $Geo_{IEI-1}$ (a) and $Geo_{IED-1}$ (b, side and top views), higher in energy than $Geo_{IEI-1}$ by 9 kJ/mol. The centre of charge is indicated with a white circle.}
        \label{fig:geo_BzW}
    \end{figure}

Before studying clusters involving several water molecules, we decided to benchmark methods and basis on Bz for which detailed experimental results exist and to investigate the effect of the coordination of one water molecule on the electronic spectrum of Bz considering two orientations (see Fig.~\ref{fig:geo_BzW}):  $Geo_{IEI-1}$ in which one hydrogen atom of the water molecule interacts with the $\pi$ cloud of benzene, leading to an increase of the vertical ionisation energy (VIE) of +0.32\,eV, and $Geo_{IED-1}$, in which the oxygen atom interacts with two hydrogen atoms of Bz, leading to a decrease of the VIE by 0.33\,eV. These VIE values were obtained using the charge constraint SCC-DFTB scheme \cite{simon_dissociation_2017}, quoted hereafter C-DFTB, previously used to compute the VIEs of large PAH-water clusters systems \cite{michoulier_theoretical_2018}. Experimental geometry was used for Bz \cite{1966msms.book.....H} while the geometries of the two systems with one water molecule were optimized using the same C-DFTB potential as for the other clusters studied in this work for the sake of consistency. \\

The structures of the $(C_6H_6)(H_2O)_n$ (n $>$ 1) clusters were extracted from larger systems containing more than 100 water molecules organized as Ih ice and optimized at the SCC-DFTB (quoted hereafter DFTB) level of theory \cite{michoulier_perturbation_2020,michoulier_theoretical_2018}. 
All configurations differ by the local organisation at the Bz-ice interface \cite{michoulier_theoretical_2018}. We chose to study two series of systems:  one for which the  adsorption on ice leads to the maximum increase of the ionisation energy ("$Geo_{IEI-n}$" series) and, on the opposite, one for which the  adsorption on ice leads to the maximum decrease of the ionisation energy ("$Geo_{IED-n}$" series).  \\

\begin{figure}
    \centering
    \subfigure[$Geo_{IEI-n}$, top view]
    {
        \includegraphics[width=0.45\textwidth]{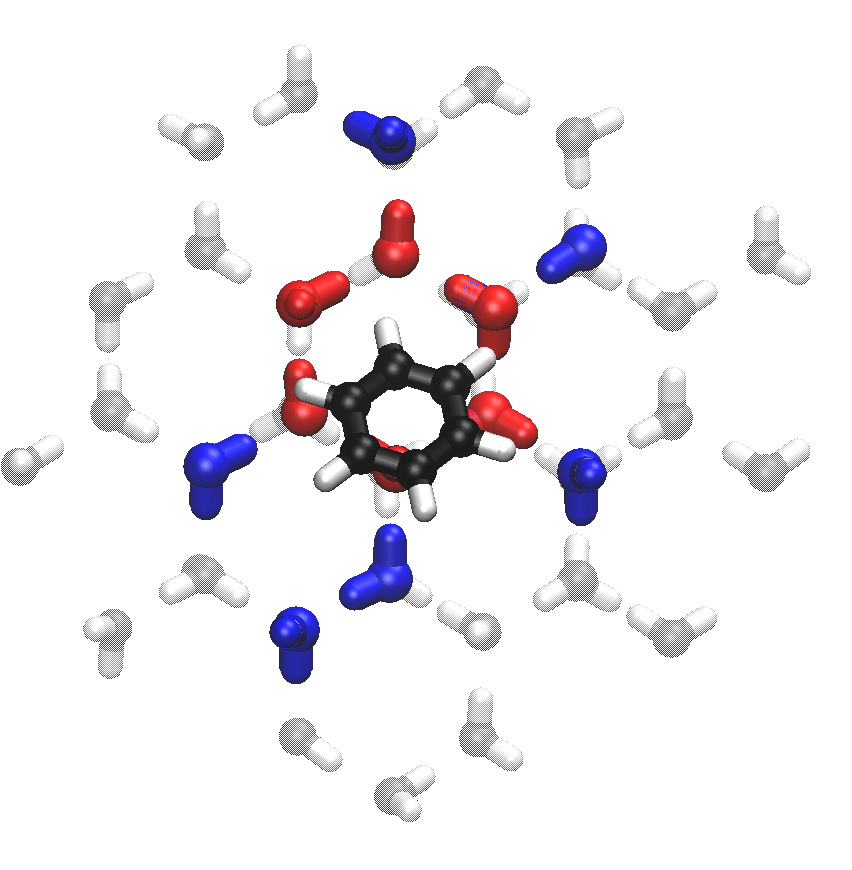}
        \label{fig:geom_IPI_IPD_1_sub}
    }
    \subfigure[$Geo_{IED-n}$, top view]
    {
        \includegraphics[width=0.45 \textwidth]{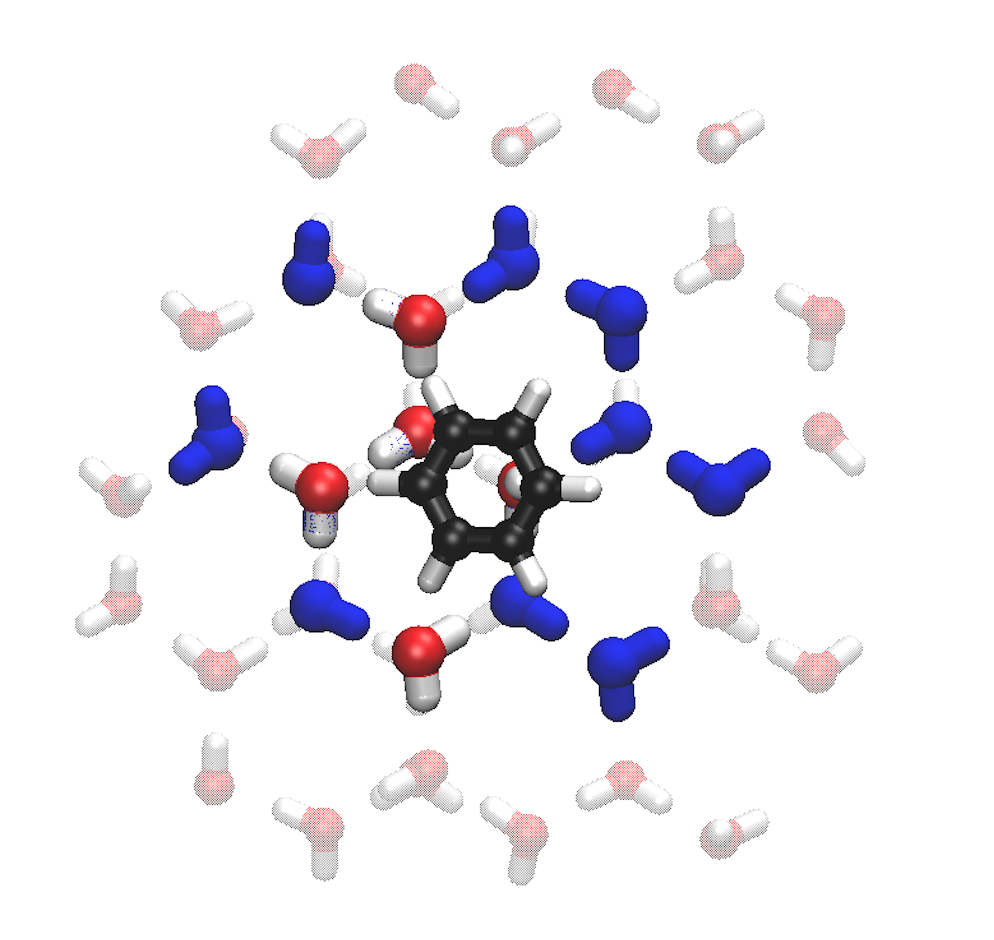}
        \label{fig:geom_IPI_IPD_2_sub}
    } \\
    \subfigure[$Geo_{IEI-n}$, side view]
    {
        \includegraphics[width=0.45 \textwidth]{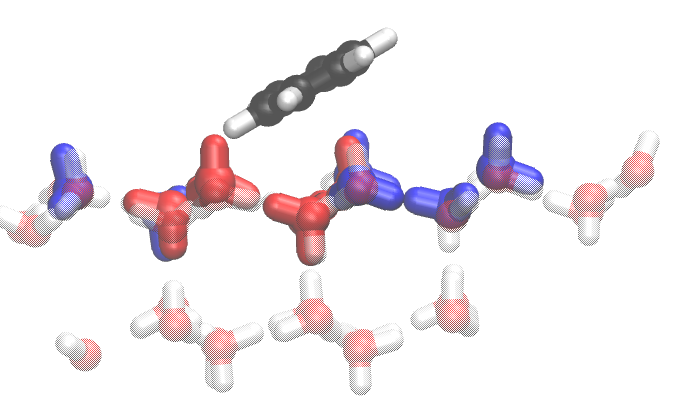}
        \label{fig:geom_IPI_IPD_3_sub}
    }
    \subfigure[$Geo_{IED-n}$, side view]
    {
        \includegraphics[width=0.45 \textwidth]{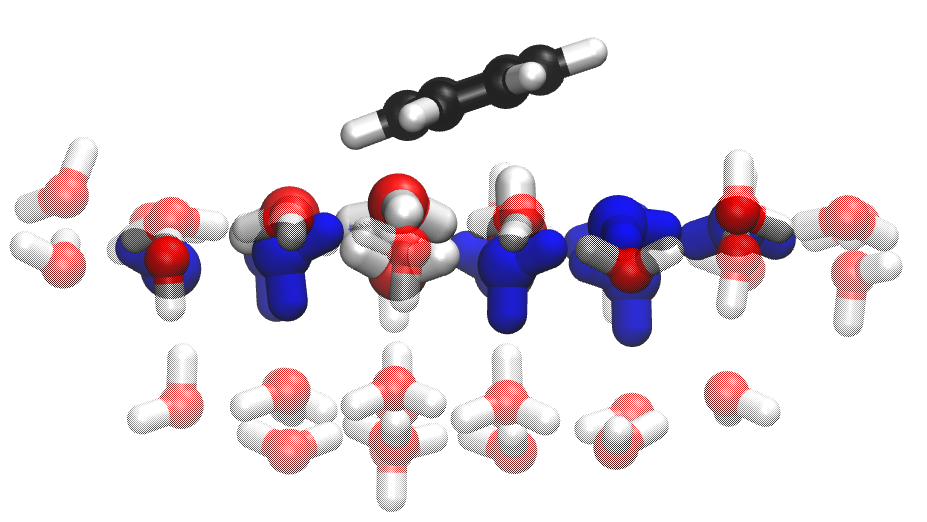}
        \label{fig:geom_IPI_IPD_4_sub}
    }
    \caption{Configurations leading to an increase/decrease of the vertical ionisation energy of Bz (IEI/IED). Red water molecules correspond to the smallest systems (n$_{H_2O}$=5,6), blue water molecules are the additional water molecules for n$_{H_2O}$=12,14 while the transparent water molecules complete the set for n$_{H_2O}$=48,49 (see text).}
    \label{fig:geom_IPI_IPD}
\end{figure}

In order to achieve electronic excited states calculations maintaining explicit electronic structure for all atoms, we had to reduce the size of the system. The truncations were achieved following the procedure detailed hereafter. Shells of decreasing radius were determined as follows: for a given water molecule, we
defined the metric d(C–O) as the minimum distance between
its oxygen atom and any of the carbon atoms of the Bz molecule.
A water molecule was kept if
d(C–O) was  smaller than 8.0\,\AA\,($Geo_{IEI-49}$ and $Geo_{IED-48}$), 5.0\,\AA\, ($Geo_{IEI-12}$ and $Geo_{IED-14}$) and 3.5\,\AA\,($Geo_{IEI-6}$ and $Geo_{IED-5}$).  The $Geo_{IEI-n}$ and $Geo_{IED-n}$ geometries studied in this work are reported in Fig.~\ref{fig:geom_IPI_IPD}. The VIEs of the solvated Bz molecule was estimated using the C-DFTB approach as in \cite{michoulier_theoretical_2018}. 
We used the same parameterisation as in our previous work \cite{michoulier_theoretical_2018}. We must note that, with such parameterisation, the VIE of Bz is overestimated by 0.16\,eV with respect to the experimental value, although in the case of pyrene, it was found equal to the experimental ionisation energy \cite{michoulier_theoretical_2018}.  
All the geometries' cartesian coordinates are given in the Supplementary Information (SI).
\newfloat{Figure}{H}{lof}

\subsection{Electronic structure methods and basis sets}\label{subsec:compdct}


Complete Active Space Self Consistent Field (CASSCF) \cite{roos_complete_1980}  and MS-CASPT2 \cite{finley_multi-state_1998} methods have been used for the determination of the excitation energies of the smallest systems in order to get reference data to benchmark TD-DFT calculations when Rydberg or CT states are present. The CASSCF/MS-CASPT2 calculations were carried out with the 7.8 MOLCAS package \cite{Molcas,Molcas2,Molcas3}. 
Atomic Natural Orbitals (ANO-L) \cite{widmark_90} basis sets were used with the following contraction scheme: for C, N, O a (14s9p4d3f) set contracted to [3s2p1d] and for H a (8s4p3d) set contracted to [2s1p] were used. As Rydberg states are interleaved with valence states in the Bz molecule, specific diffuse functions with small Gaussian exponents, in the average position of the two centers of charge of the cations coming from the two highest $\pi$ orbitals of Bz, were defined for each studied system \cite{carsky_ab_1980}. 
The additional diffuse functions were obtained using exponents from the Kaufmann series \cite{Kaufmann_1989} and coefficients obtained from the Genano module in MOLCAS \cite{lorentzon_caspt2_1995}. All the coefficients were kept, as a very large number of Rydberg states was expected, not only those of the Bz molecule but also those of the increasing number of water molecules. Furthermore, all the states of isolated Bz are better described (see SI, Table S1), even the valence states, while the Rydberg states are considerably affected. These coefficients were determined for each system and can be found in the SI. These basis sets are called Gen. in the tables.\\
 The active space was defined to provide a good description of valence singlet states ($\pi$ $\rightarrow$ $\pi*$) as well as the Rydberg states ($\pi$ $\rightarrow$ $Ry$). The same active space was defined for all systems, with all $\pi$ orbitals of Bz and one s, three p and five d Rydberg orbitals. However, including oxygen lone pairs in the active space was not feasible and the $n_O$ $\rightarrow$ $R_y$ states could not be described. State-average orbitals on 42 singlet states were optimized in order to obtain all the targeted states. \\

At the MS-CASPT2 level, the 34 lowest states were considered. The use of a level shift in the MS-CASPT2 method avoids weak intruder states by the addition of a shift parameter to the zeroth-order Hamiltonian. The value of this level shift was chosen to obtain a stability of the excitation energies and a maximal deviation of 3\% of the reference weight of the zeroth-order wave functions of the excited states compared to the ground state, {\it ie} 0.3 for Bz and 0.4 for all other systems. Standard ionization potential-electron affinity (IPEA) shift in the perturbation treatment was used as MS-CASPT2 underestimates excitation energies. The oscillator strengths were calculated thanks to the restricted active space state interaction approach (RASSI) \cite{malmqvist_rassi_2002}.\\ 

 The electronic spectra were also computed using the time dependent DFT methodology using the CAM-B3LYP functional \cite{camb3lyp}, a long range corrected functional that was shown to provide satisfactory results for the description of low lying Rydberg states of PAHs \cite{bohl_low-lying_2017} as well as Rydberg and CT intra- and intermolecular states for a series of test molecules \cite{maier_validation_2016}.
  Two different basis sets were tested on isolated Bz for benchmark. The first basis is the same as for the MS-CASPT2 calculation, named Gen., and the second basis is a "SVP + diffuse" basis \cite{dunning_gaussian_1977} with  diffuse functions centered at the average position of the centers of charge of the two cations obtained by the removal of one electron from Bz. These functions are those of the "DZ + double Rydberg" defined for the carbon atom (2s2p2d) \cite{dunning_gaussian_1977} with one additional f function (coefficient of 0.012). For all the other systems (Bz with water molecules), the basis set to describe the Rydberg orbitals is constructed by averaging the coefficients of the carbon and oxygen atoms. 
 This second basis set is called DunRy (2s2p2d1f) in the manuscript. \\
 
 In order to characterize the CT character of the electronic excited states and in particular to investigate the presence of spurious states (ghost states), we used descriptors namely the D$_{\text{CT}}$ index \cite{le_bahers_qualitative_2011} as implemented in Gaussian16 and the related M$_{\text{AC}}$ (Mulliken average configuration) index \cite{campetella_charge_2017}. The D$_{\text{CT}}$ index quantifies the spatial extent associated to an electronic transition, hence it should provide indication on the CT character of the transition. For a given transition, the comparison between the M$_{\text{AC}}$ value and the energy should allow us to get insights into the potential ghost character of the excited states. However, they have never been used to characterize the type of systems studied in the present paper, but rather on extended delocalised systems. The D$_{\text{CT}}$  and M$_{\text{AC}}$ values  were computed for all systems for the unrelaxed densities. When ghost states were suspected, these were re-evaluated with the excited states relaxed densities to insure quantitative results \cite{maschietto_how_2018}. \\
 
 The TD-DFT calculations and their related properties, natural transitions orbitals (NTOs) \cite{NTO_2003} and D$_{\text{CT}}$ index, were computed using the Gaussian16 suite of programs \cite{g16}. The TD-DFT energies and excitations' detailed assignments, the shape of the molecular orbitals (MOs), a list of the transitions with the corresponding D$_{\text{CT}}$ and M$_{\text{AC}}$ index, as well as the NTOs for Geo$_{IEI-12}$ and  Geo$_{IEI-49}$ can be found in the .pdf files of the SI (see section~\ref{sec:SI}). 
All calculations have been carried out in the C$_1$ point-group symmetry. \\
 
 \subsection{Benchmark on benzene}
  

We chose to use the geometry of the Bz molecule taken from the Hertzberg book \cite{1966msms.book.....H}. Benchmark was achieved on Bz using experimental results. The experimental spectroscopic data arise from absorption spectra and multi-photon measurements \cite{pantos_extinction_1978,nakashima_laser_1980,whetten_higher_1985,johnson_discovery_1983,grubb_higher_1985,hiraya_direct_1991}. The Bz molecule has also been extensively studied by various theoretical methods such as Coupled Cluster (CC) \cite{christiansen_largescale_1996,del_bene_coupled-cluster_1997,nooijen_similarity_1999,schreiber_benchmarks_2008,falden_benchmarking_2009,silva-junior_basis_2010,li_multi-reference_2011,kannar_benchmarking_2014,loos_mountaineering_2018,dutta_exploring_2018,loos_reference_2019}, Perturbation theory \cite{lorentzon_caspt2_1995,schreiber_benchmarks_2008,loos_mountaineering_2018,loos_reference_2019,roos_towards_1992,hashimoto_theoretical_1998,devarajan_generalized_2008,silva-junior_benchmarks_2010,sauri_multiconfigurational_2011,sharma_valence_2019}, Multirefence Configuration Interaction (MRCI) \cite{palmer_electronic_1989,sobolewski_ab_1993,loos_reference_2019},  Symmetry Adapted Cluster—Configuration Interaction \cite{li_theoretical_2007} or
Time Dependent Density Functional \cite{jacquemin_absorption_2006,leang_benchmarking_2012,schwabe_time-dependent_2017,sharma_valence_2019}. In most of these studies, only low-lying states ($\pi$  $\rightarrow$ $\pi^\star$ and $\pi$  $\rightarrow$ $Ry$ 3s,3p) were investigated. 
In some cases \cite{lorentzon_caspt2_1995,christiansen_largescale_1996,hashimoto_theoretical_1998,li_theoretical_2007,falden_benchmarking_2009} 3d Rydberg states were also included and the MRCI calculations by Palmer and Walker \cite{palmer_electronic_1989} include Rydberg states up to n=4. In the present work, the MS-CASPT2 wave functions and energies were also analysed and compared to Coupled Cluster including triple excitations EOM-CCSDT when available or to CC3 which is an approximation  of the latter.\\

{\bf MS-CASPT2 calculations.} The first issue concerns the generation of the ANO basis set for Rydberg states. As mentioned before, when only one contracted function is used for each angular momentum component, the MS-CASPT2 error compared to experimental data is considerably larger than when keeping a larger set of functions (3s3p4d orbitals).
 Indeed, the excitation energies are reduced by 0.1-0.14\,eV for valence states, 0.06-0.08\,eV for the Rydberg 3s states, $\sim$ 0.4\,eV for the Rydberg 3p and 0.5-1.08\,eV for the Rydberg 3d states (Table S1 in SI).

The MS-CASPT2 $\pi$$\pi^*$ valence states are in good agreement with the experimental values ($<$0.1 eV) except the $^1E_{2g}$  state ($\pi$ a$_{2u}$ $\rightarrow$ $\pi^\star$) which is overestimated by 0.3 eV. For the first excited state ($^1B_{2u}$) the weight of single excitations was found to be 79\% with 17\% of the transition possessing double excitation character. For the two following valence states ($^1B_{1u}$  and $^1E_{1u}$) the weights of single excitations were found 93\% and 85\%  respectively while the $^1E_{2g}$  state, lying at 8.08\,eV, was found to be composed of 45\% of single excitations (32\% of $\pi$ a$_{2u}$ $\rightarrow$ $\pi^\star$) and 45\% of double excitations. The Rydberg states consist in single excitations for 88\% - 89 \% except the $^1A_{2u}$  state at 9.28\,eV which corresponds mainly to the $\pi$ a$_{2u}$ $\rightarrow$ $Ry$ (3s) transition (61\%) with 29\% of double excitations. The $^1E_{1g}$  ($\pi$ $\rightarrow$ $Ry$ 3s), $^1A_{2u}$, $^1E_{2u}$, $^1A_{1u}$   ($\pi$ $\rightarrow$ $Ry$ ($3p_x$,$3p_y$)) and $^1A_{1g}$, $^1E_{2g}$ , $^1A_{2g}$   ($\pi$ $\rightarrow$ $Ry$ (3d$_{xz}$,3$_{dyz}$)) states differ from the experimental values by almost 0.1\,eV while for the $^1E_{1g}$  ($\pi$ $\rightarrow$ $Ry$ 3d$_{z^2}$) transition, the discrepancy is only of 0.03\,eV. The $^1E_{1u}$ ($\pi$ $\rightarrow$ $Ry$ $3p_z$) and $^1B_{1g}$, $^1B_{2g}$, $^1E_{2g}$($\pi$ $\rightarrow$ $Ry$ (d$_{x^{2}}$$_{-y^{2}}$, 3d$_{xy}$)) were found +0.2 eV higher.\\
The MS-CASPT2 values computed in this work show small differences with the few available CCSDT ones (0.09\,eV) \cite{loos_mountaineering_2018} and the energy gaps between excited states are almost identical except for the $^1B_{1u}$  state. At the MS-CASPT2 level, this state was found in better agreement with the experimental value (no difference). Eventhough the CC3 method is an approximation of the CCSDT method, there is usually little difference between the results. In Table~\ref{tbl:Bz}, the difference between the two last columns 
arises from the use of different basis sets. As mentioned by Christiansen {\it et al.} \cite{christiansen_largescale_1996}, the Rydberg states of Bz are better described at the CCSD level (about 0.1 eV) by more flexible basis sets such as aug-cc-pVTZ {\it vs} aug-cc-pVDZ, both supplemented with basis set functions at the center of charge. As more states have been calculated for Bz at the CC3 level, the comparison of MS-CASPT2 calculations was also done with this method. The overestimation of the MS-CASPT2 values compared to the experimental ones was found smaller than the CC3 \cite{christiansen_largescale_1996} value for the valence states and comparable for the Rydberg states. Coupled cluster methods including triple excitations are expected to describe very accurately pure single excited states and less accurately states with large amount of double excitation character such as the valence $^1E_{2g}$ state (32\% $\pi$ a$_{2u}$ $\rightarrow$ $\pi^\star$ and 45\% double excitations) for which CC3 found an excitation energy 0.6\,eV higher than the experimental value \cite{christiansen_largescale_1996}. However, the comparison between vertical transitions computed using state-of-the-art {\it ab initio} methods  and  experimental 0-0 origin transitions is not straightforward as geometry relaxation and zero-point vibration should be taken into account.  
To conclude, one can note the non negligible role of the basis sets, especially on the Rydberg states, although this has been taken into account by adding functions to the charge center of the molecule. Furthermore, the results obtained at the MS-CASPT2 level were found reliable enough to be used for the present work's goal which is to investigate the effect of adsorption of water molecules and clusters on the electronic spectrum involving Rydberg and charge transfer states of larger systems. \\
\\
{\bf TD-DFT calculations.} The excited states were then determined with these basis sets using TD-DFT with the CAM-B3LYP functional. Almost all the states were found with a maximum error of 0.12\,eV except i) the first valence $\pi$ $\rightarrow$ $\pi^\star$ excited state which appears badly described (error of 0.55\,eV compared to experimental value), this state presents a non-negligible double excitation character (17\% at the MS-CASPT2 level); ii) as expected, the two upper states with large amount of double excitation character are badly reproduced by TD-DFT method, 9.48\,eV for the valence state and 9.99\,eV for the Rydberg state. There were found so high in energy (states 67, 68 and 88) that many other excited states appear at lower energy, mainly Rydberg n=4 states which were not reported in Table 1. There was an admixture of valence and Rydberg states at 7.04  ($^1E_{1u}$)  and 7.31\,eV ($^1E_{1u}$ (3p$_z$)). The corresponding oscillator strengths are shared and lead to two absorption bands instead of one. The CASSCF method found this ionic valence state ($^1E_{1u}$) too high in energy, close to another Rydberg state ($\pi$ $\rightarrow$ $\mathrm{3d (d_{xz}, d_{yz})}$), leading to valence-Rydberg mixing which disappears thanks to the MS-CASPT2 method. 

Next issue concerns the Rydberg basis sets when using TD-DFT method. As segmented and smaller basis set is preferable and even necessary to study the largest systems using Gaussian16, the DunRy (2s2p2d1f) basis set was compared to the previous one. The differences with experimental excitation energies were found to be small, except for the first valence state, as previously, leading to a maximum deviation to the experimental value of 0.10\,eV for all the other states. A similar admixture of valence/Rydberg states was also found,  to a lesser extent as with the Gen. basis set, as the $\pi$ $\rightarrow$ $\pi^\star$ was found at 6.99\,eV (DunRy) instead of 7.04\,eV (Gen.) with oscillator strengths of 0.49 {\it vs} 0.44. Similarly, the $\pi$ $\rightarrow$ $Ry$ (3p$_z$) was found at 7.27\,eV (DunRy) instead of 7.31 (Gen.) with oscillator strengths of 0.14 {\it vs} 0.18. 
All the methods and basis sets give accurate VIEs with respect to experimental data, with a maximum deviation of 0.02 eV. 

Finally, the TD-DFT using the CAM-B3LYP functional with "SVP+ diffuse" Dunning-Hay basis set for Bz adding diffuse 2s2p2d1f functions to describe the Rydberg states, provides very good results for Bz compared to experimental values, except for the first valence excited state, which however has no oscillator strength and is not significant in the study of the effect of solvation on the electronic states, and in the higher part of the spectrum ($^1E_{2g}$  state at 7.8 eV). In view of these results, the following TD-DFT calculations are performed within these basis sets.\\

\newfloat{Table}{H}{lot}
\begin{table}
\begin{minipage}[t]{15cm}
\renewcommand{\footnoterule}{} 
\begin{tabular}{|ll|l|l|l|l|l|l|}
\hline
\multicolumn{2}{|c|}{Transition} & MS-CASPT2/ & \multicolumn{2}{c|}{TD-DFT/} & Exp. \footnote { \cite{pantos_extinction_1978,nakashima_laser_1980,whetten_higher_1985,johnson_discovery_1983,grubb_higher_1985,hiraya_direct_1991}} & CC3\footnote{CC3 results from Christiansen {\it et al.} \cite{christiansen_largescale_1996} with ANO \cite{widmark_90} basis set: C 4s3p1d, H: 2s1p and basis set at center of charge contracted to 1s1p1d} & CC3\footnote{CC3/CCSDT results from Loos {\it et al.} \cite{loos_mountaineering_2020} with aug-cc-pVTZ basis set} \\
Nature & Symmetry & Gen. & Gen. & DunRy (2s2p2d1f)  & & &(CCSDT)\\
\hline
\multicolumn{8}{c}{Valence states}\\
\hline
$\pi$ $\rightarrow$ $\pi^\star$                 & $^1B_{2u}$ & 4.97       & 5.45    & 5.46 & 4.90 & 5.08 & 5.09 (5.06) \\
                                                & $^1B_{1u}$ & 6.20       & 6.17    & 6.16 & 6.20 & 6.54 & 6.44 (6.45) \\
                                                & $^1E_{1u}$ & 6.91 (.80) & 7.04 (.44) & 6.99 (.49) VR & 6.94 & 7.13&\\
                                                &            & 6.91 (.80) & 7.04 (.44) & 6.99 (.49) VR & &&\\
$\sigma$ $\rightarrow$ $\pi^\star$  &           & & 7.91 & 7.92 & &&\\
                                    &           & & 8.01 (.009) & 8.02 (.007) &&& \\
                                    &           & & 8.02 & 8.03 &&& \\ 
                                    &           & & 8.02 & 8.03 &&&\\
$\pi$ $(a_{2u})$ $\rightarrow$ $\pi^\star$      & $^1E_{2g}$ & 8.08       & 9.48       & 9.43          & 7.8& 8.41&\\
                                                &            & 8.10       & 9.48       & 9.44          &  & &\\
\hline
\multicolumn{8}{c}{Rydberg states n=3}\\
\hline
$\pi$  $\rightarrow$ $\mathrm{3s}$               & $^1E_{1g}$ & 6.43 &6.38 & 6.38 & 6.33 & 6.51 & 6.52 (6.52)\\
                                                 &            & 6.44  &6.38  &6.38& &&\\
$\pi$ $(a_{2u})$ $\rightarrow$ $\mathrm{3s}$     & $^1A_{2u}$ & 9.28 (0.08) & 10.10 (.05)& 9.99 (.12)& &&\\
                                                 &            &            &&&&&\\
$\pi$ $\rightarrow$ $\mathrm{3p (px,py)}$        & $^1A_{2u}$ & 7.00 (.10) &6.89 (.07) &6.84 (.06) &6.93 & 6.97 & 7.08 (7.08) \\
                                                 & $^1E_{2u}$ & 7.07 (.002) &6.99 & 6.94 &6.95 & 7.03 & 7.15 (7.15) \\
                                                 &            & 7.07 (.001) &6.99 & 6.94 && & \\
                                                 & $^1A_{1u}$ & 7.14 & 7.10 &7.05& &7.11&\\
                                                 &            &            &&&&&\\
$\pi$ $\rightarrow$ $\mathrm{3p (pz)}$           & $^1E_{1u}$ & 7.21 (.06) &7.31 (.18) & 7.27 (.14) VR & 7.41 & 7.42&\\
                                                 &            & 7.22 (.05) & 7.31 (.18) & 7.27 (.14) VR&&&  \\
                                                 &            &            &&&&&\\
$\pi$ $\rightarrow$ $\mathrm{3d (dz^2)}$         & $^1E_{1g}$ & 7.57 & 7.48 & 7.47 & 7.54 & 7.69&\\
                                                 &            & 7.57 & 7.48 & 7.47 && &\\
                                                 &            &            &&&&&\\
$\pi$ $\rightarrow$ $\mathrm{3d(dx^2-y^2, dxy)}$ & $^1B_{1g}$ & 7.66 & 7.57 &7.52 & 7.46 & 7.65 &\\
                                                 & $^1B_{2g}$ & 7.67 & 7.57 & 7.53 &     & 7.65 &\\
                                               &  $^1E_{1g}$ & 7.73 & 7.58 & 7.55 &     & 7.63 &\\
                                                &  & 7.73 & 7.58 & 7.55 &     &  &\\
                                         
                                                 &            &            &&&&&\\
$\pi$ $\rightarrow$ $\mathrm{3d (dxz, dyz)}$     & $^1A_{1g}$ & 7.89 & 7.78 & 7.70 & 7.81 & 7.86&\\
                                                 & $^1E_{2g}$ & 7.91 & 7.78 & 7.71 & 7.81 & 7.85&\\
                                                 &  & 7.91 & 7.78 & 7.71 &&& \\
                                                 &          $^1A_{2g}$  & 7.94 & 7.83 & 7.75 && 7.88&\\

\hline
\hline
VIE && 9.26 &9.25 &9.26 (DFT)& 9.24 &&\\
 &  &&& 9.40 (DFTB) && &\\
\hline
\end{tabular}
\end{minipage}
\caption{MS-CASPT2/CAS(6,15) (DZP+Genano basis sets) / TD-DFT (CAM-B3LYP with Dunning-Hay Rydberg or Genano basis sets) electronic excitation energies (in eV) for $\mathrm{C_6H_6}$ (Bz). 
 All excitations involve the $\pi$ ($1e_{1g})$ orbitals unless it is specified that the $\pi$ ($a_{2u}$) orbital is involved. Vertical ionisation energies (VIEs) were determined using CAS(5,6)PT2 for two cations and the neutral ground state obtained at the CAS(6,6)PT2 level of theory (CAS restricted to $\pi$ orbitals). 
 In the last three columns are given reference values, experimental data ("Exp.") and theoretical CC3 and CCSDT results. } 
\label{tbl:Bz}

\end{table}

\section{Results}\label{sec:res}

In this section, we present the electronic spectra of $Geo_{IEI-n}$ and $Geo_{IED-n}$  computed at the TD-DFT (CAM-B3LYP, n $<$ 50) and MS-CASPT2 (n $\leq$ 6) levels using the basis sets determined as explained hereabove. We chose to report in the main article the electronic spectra, the assignments of the transitions and some NTOs of interest.  The details of the transitions that were found to be extremely multideterminental, and the MOs are reported in the SI, along with the $D_{\text{CT}}$ and $M_{\text{AC}}$ indexes for all transitions of all systems. Only transitions with weights larger than 0.04 were taken into account to assign the nature of the transitions. The nature of the excited states is discussed in the light of these analysis.  
All electronic spectra result from the convolution of the discrete spectra by a Lorentzian profile with a full width at half maximum of 1.7 10$^{-3}$\,eV. \\ 

As we compare MS-CASPT2 and TD-DFT results, we must specify that, by construction, transitions from $n_O$ orbitals are not described at the MS-CASPT2 level whereas they can be observed by essence at the TD-DFT level. Besides, regarding the MOs, and for all systems studied in the present work, a general observation is that more mixing between the different Rydberg orbitals, and also more valence-Rydberg mixing were found in TD-DFT calculations that in MS-CASPT2 calculations. However, local orbitals have been used in this latter case, which simplifies considerably the analysis by avoiding the mixing of the orbitals.\\

\subsection{$Geo_{IEI-1}$ and $Geo_{IED-1}$}

\begin{figure}
        \centering
        \includegraphics[width=14cm]{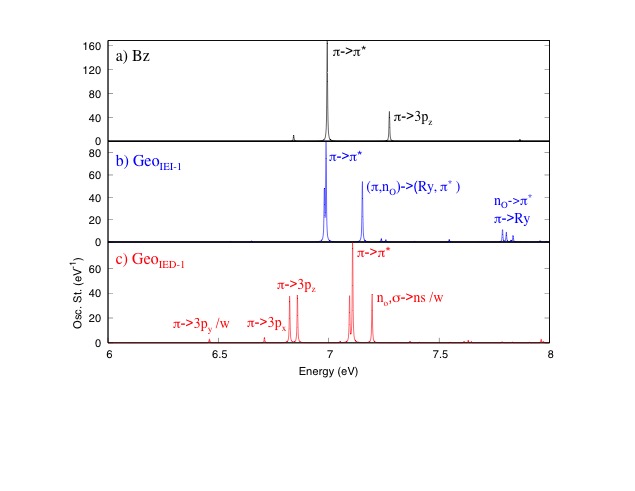} \\
          \vspace{-2.5cm}
         \includegraphics[width=14cm]{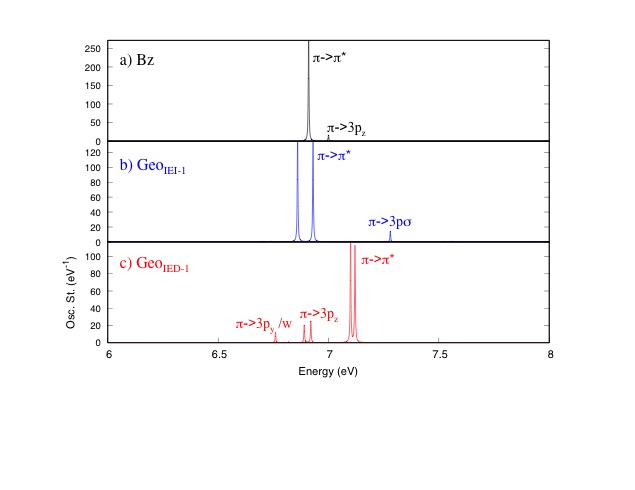} \\
        \vspace{-2.5cm}
        \caption{Electronic spectra computed at the TD-DFT (top) and MS-CASPT2 (bottom) levels of theory 
        and assignement of most intense transitions for a) Bz (black), b) and c) : $Bz-H_2O$ in $Geo_{IEI-1}$ (b, blue) and $Geo_{IED-1}$ (c, red)}
        \label{fig:spec}
    \end{figure}

\begin{figure}
        \centering
        \includegraphics[width=8cm]{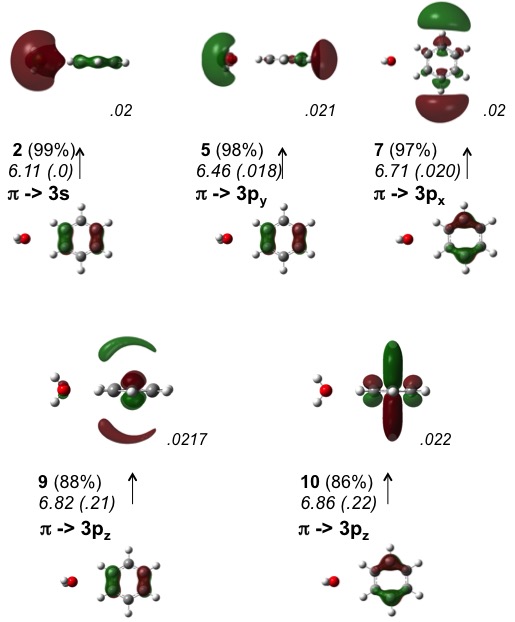}
        \caption{NTOs involved in the lowest energy and most intense transitions in $Geo_{IED-1}$ (see (c) in Fig.~\ref{fig:spec} and assignment of transitions in table \ref{tbl:Bz-H2O_geo3}). The isovalues chosen for the LUMOs are specified in italics, a value of 0.08 was chosen for the HOMOs. Are also reported : the number of the transition, the contribution of the transition described by the NTOs with respect to the complete excited state (in \%). The energy of the transition (in eV) and the oscillator strength are reminded in italics, the latter being in parenthesis. }
        \label{fig:nto_geo3}
    \end{figure}

The TD-DFT and MS-CASPT2 electronic spectra of $Geo_{IEI-1}$ and $Geo_{IED-1}$ along with that of Bz, are reported in Fig.~\ref{fig:spec}. The assignments of the most intense bands are reminded. The analysis of the transitions (only the lowest energy ones of the TD-DFT spectra) are analysed in Table \ref{tbl:Bz-H2O_geo1} for $Geo_{IEI-1}$ and in Table \ref{tbl:Bz-H2O_geo3} for $Geo_{IED-1}$, based on the transitions and MOs reported in the SI (Geo\_IEI-1.pdf and Geo\_IED-1.pdf files).  The NTOs corresponding to the lowest energy transitions  describing electron transfer from $\pi$ to Rydberg orbitals, mixed with diffuse MOs for $Geo_{IED-1}$, are reported in 
Fig.~\ref{fig:nto_geo3}.\\

\newfloat{Figure}{H}{lof}

At both levels of theory, we see that the effect of the adsorption of the water molecule on the electronic spectrum of Bz depends on their relative orientations. The shifts of the transitions within Bz induced by the adsorption of the water molecule are reported in Tables~\ref{tbl:Bz-H2O_geo1} and~\ref{tbl:Bz-H2O_geo3}  for $Geo_{IEI-1}$ and $Geo_{IED-1}$ respectively ($\delta$E column). For instance, the $\pi \rightarrow$ Ry transitions are blue-shifted in the case of $Geo_{IEI-1}$ and red-shifted in the case of $Geo_{IED-1}$.
In $Geo_{IEI-1}$, the position of the intense $\pi$ $\rightarrow$ $\pi^\star$ transition  is hardly affected by the interaction with the water molecule. A splitting due to symmetry breaking is observed. The difference between the TD-DFT and MS-CASPT2 levels is that this splitting  is slightly larger at the MS-CASPT2 level (6.98/6.99\,eV {\it vs} 6.86/6.93\,eV). The other intense transitions at 7.15\,eV at the TD-DFT level are significantly mixed transitions (see "Mixed states" columns in Table~\ref{tbl:Bz-H2O_geo1})  
whereas the quite intense MS-CASPT2 state, observed at 7.28\,eV, was found to be a pure $\pi \rightarrow$ 3p$_{\sigma}$ state.

In $Geo_{IED-1}$, the intense $\pi$ $\rightarrow$ $\pi^\star$ transitions are split and shifted towards higher energy at both MS-CASPT2 and TD-DFT levels. The blue-shift is larger at the MS-CASPT2 level (7.10/7.12\,eV {\it vs} 6.91\,eV for Bz) than at the TD-DFT level (7.09/7.11\,eV {\it vs} 6.99\,eV for Bz). The splitting was found the same for both methods. 
Interestingly, contrary to $Geo_{IEI-1}$, low energy transitions  corresponding to a promotion of an electron from a $\pi$ orbital to a 3p (and 3s for TD-DFT) Rydberg orbital located on the water molecule are found at both levels of theory, with non negligible intensities. More precisely,  a state involving mainly a $\pi \rightarrow 3p_y /{\it w}$ transition was found at 6.76\,eV at the MS-CASPT2 level and at 6.46\,eV at the TD-DFT level (see NTO 5 on Fig. \ref{fig:nto_geo3}).
Besides, an intense transition mainly from the $n_O$ orbital of the water molecule  
to a Rydberg orbital centered on the water molecule is seen at 7.20\,eV (see SI, Geo\_IED-1.pdf file). Removing the Bz molecule, such  $n_O$ $\rightarrow$ $Ry$ (3s) was found at 7.03\,eV (.06) for the water molecule in the same geometry as in $Geo_{IED-1}$. For $Geo_{IEI-1}$ there is only a small contribution of the $n_O$  $\rightarrow$ $Ry$ excitation in the $\pi$ $\rightarrow$ $\pi^\star$ state at 6.98 eV. The corresponding Rydberg state in the water molecule at the same geometry was found at 6.99\,eV (.05).\\
Interestingly, the $\pi$ $(a_{2u})$ $\rightarrow$ $\pi^\star$ Bz transitions are hardly affected by the coordination of the water molecule and their energies remain overestimated by more than 1\,eV at the TD-DFT {\it vs} MS-CASPT2 level.  \\

\newfloat{Table}{H}{lot}
\begin{table}
\begin{tabular}{|l@{\hskip.8mm}|l@{\hskip.8mm}r@{\hskip.8mm}|l@{\hskip.8mm}r@{\hskip.8mm}|}
\hline
&  \multicolumn{4}{c|}{$Geo_{IEI-1}$}        \\

Transition & \multicolumn{2}{c|}{MS-CASPT2/Gen.} & \multicolumn{2}{c|}{TD-DFT/DunRy (2s2p2d1f)}        \\
& Energy  & $\delta$E /Bz  & Energy & $\delta$E /Bz        \\
\hline
\multicolumn{5}{|c|}{Valence states (mixed with Ryd. in TD-DFT ) }       \\
\hline

$\pi$ $\rightarrow$ $\pi^\star$                     & 4.94 (0.001)                     & -0.03  & 5.47       & +0.01 \\
                                                    & 6.19 (0.002)                     & -0.01  & 6.17       & +0.01 \\
                                                    & 6.86 (0.790)                     & -0.05  & 6.98 (.26) & -0.01 \\
                                                    & 6.93 (0.793)                     & +0.02  & 6.99 (.48) &  0.00 \\
                          $\pi_1$ $\rightarrow$ $\pi^\star$                          & 8.00 (3d$_{xy}$)                 & +0.08  & 9.17       & -0.26 \\
                                                    & 8.14                             & +0.04  &           &      \\
$n_O$ $\rightarrow$ $\pi^\star$                 &           & & 7.79 (.06)   &      \\
                                                &           & & 7.80 (.05)  &      \\
$\sigma$ $\rightarrow$ $\pi^\star$                  &                                  &        & 7.87        &  -0.05    \\
              &  &     & 7.96 (.007) &  -0.06    \\
              &  &     & 7.98        &  -0.05    \\
              &  &     & 7.99 (.002) &  -0.04    \\

\hline
\multicolumn{5}{|c|}{Rydberg states}       \\
\hline

$\pi$ $\rightarrow$ $\mathrm{3s}$                   & 6.70 (0.004)                     & +0.27  & 6.56 (.002)& +0.18    \\
                                                    & 6.74 (0.005)                     & +0.30  & 6.65 (.005)& +0.27    \\
                                                    & 9.53 (0.070)                     & +0.25  &            &      \\
                                                    &                                  &        &            &      \\
$\pi$ $\rightarrow$ $\mathrm{3p (p_x,p_y)}$         & 7.28 (0.084)                     & +0.28  &            &   \\
                                                    & 7.31                             & +0.24  & 7.24 (.017)& +0.30   \\
                                                    & 7.33                             & +0.26  & 7.26 (.012)& +0.32  \\
                                                    & 7.43                             & +0.29  & 7.39 (.002)& +0.34   \\
$\pi$ $\rightarrow$ $\mathrm{3p (p_z)}$             & 7.52 (0.003) (3p$_z$/{\it w})    & +0.31  & 7.51 ($\mathrm{3d_{xz}}$)       & +0.24 \\
                                                    & 7.56 (0.005)                     & +0.34  & 7.55 (.01) ($\mathrm{3d_{xz}}$) & +0.28 \\
                                                    &                                  &        &            &      \\
$\pi$ $\rightarrow$ $\mathrm{3d}$                   & 7.92 (.001) (3d$_{z^2}$/{\it w}) & +0.35  & 7.75 (3d$_{z^2}$,3d$_{xy}$)      & +0.28 \\
                                                    & 8.03 (.001) (3d$_{z^2}$/{\it w}) & +0.46  & 8.29 (3d$_{z^2}$+3d$_{x^2-y^2}$) & +0.82 \\
                                                    & 7.96 (3d$_{xy}$)                 & +0.30  & 7.83 (.03) (3d$_{xy}$)           & +0.31 \\
                                                    & 8.01 (.001) (3d$_{xy}$)          & +0.34  & 7.90 (3d$_{xy}$)                 & +0.37 \\
                                                    & 8.01 (3d$_{x^2-y^2}$)            & +0.28  & 7.90 (3d$_{x^2-y^2}$)            & +0.35 \\
                                                    & 8.02 (3d$_{x^2-y^2}$)            & +0.29  & 8.29 (3d$_{x^2-y^2}$+3d$_{z^2}$) & +0.34 \\
                                                    & 8.26 (3d$_{xz,yz}$)              & +0.37  & 8.04 (.001) (3d$_{xz,yz}$)       & +0.34 \\
                                                    & 8.28 (.001) (3d$_{xz,yz}$)       & +0.37  & 8.06 (.001) (3d$_{xz,yz}$)       & +0.35 \\
                                                    & 8.32 (.003) (3d$_{xz,yz}$)       & +0.41  & 8.15 (3d$_{xz,yz}$)              & +0.44 \\
                                                    & 8.32 (3d$_{xz,yz}$)              & +0.38  & 8.17 (3d$_{xz,yz}$)              & +0.42 \\
\hline
\multicolumn{5}{|c|}{Very mixed states }       \\
\hline
 $\pi$ $\rightarrow$ 3p$_y$, 4p$_y$, $\pi^\star$, 3p$_x$, 4p$_z$/n$_O$ $\rightarrow$ 3s, 4p$_z$, $\pi^\star$      &  &   & 7.15 (.16) &    \\
            id.                             &  &        & 7.15 (.16)& \\
 $\pi$ $\rightarrow$ (4p + 3d)/n$_O$ $\rightarrow$ $\pi^\star$  &  &        & 7.69           &      \\
 $\pi$ $\rightarrow$ 4p$_z$ + 3d$_{x^2-y^2}$ + 4d/ n$_O$ $\rightarrow$ $\pi^\star$ & &  & 7.83 (.03)& \\

    \hline
VIE & 9.59 &  & 9.59 (DFT)   &\\  
    &      &  & 9.72 (C-DFTB)&\\
 \hline
\end{tabular}

\caption{MS-CASPT2 and TD-DFT electronic excitation energies (in eV) 
for $Geo_{IEI-1}$. The oscillator strengths are reported in parenthesis. The shifts of the transitions with respect to those in Bz are also reported ($\delta$E, in eV). When the orbitals of the excited states have some contributions on the water molecule or at its vicinity, it is indicated by the /{\it w} symbol. The $\pi_1$ orbital refers to the $\pi (a_{2u})$ orbital in Bz. Vertical ionization energy (VIE) values are also reported. 
}
\label{tbl:Bz-H2O_geo1}
\end{table}

\newfloat{Table}{H}{lot}
\begin{table}
\begin{tabular}{|l@{\hskip.8mm}|l@{\hskip.8mm}r@{\hskip.8mm}|l@{\hskip.8mm}r@{\hskip.8mm}|}
\hline
&  \multicolumn{4}{c|}{$Geo_{IED-1}$} \\

Transition & \multicolumn{2}{c|}{MS-CASPT2/Gen.} & \multicolumn{2}{c|}{TD-DFT/DunRy (2s2p2d1f)} \\
& Energy  & $\delta$E /Bz  & Energy & $\delta$E /Bz        \\
\hline
\multicolumn{5}{|c|}{Valence states (mixed with Ryd. in TD-DFT ) }\\
\hline
$\pi$ $\rightarrow$ $\pi^\star$ 
& 4.97 &+0.0 & 5.45 &-0.01 \\
& 6.17 (.003) & -0.03& 6.14 &+0.02\\
&7.10 (.681) &+0.19 & 7.09 (.21) (3p$_z$+3d$_{yz}$) & +0.10\\ 
&7.12 (.657)& +0.21 & 7.11 (.46) (3p$_z$+3d$_{yz}$) & +0.12  \\
$\pi_1$ $\rightarrow$ $\pi^\star$ & 8.13 (0.10) & +0.05 &9.39 & -0.04\\
& 8.13 & +0.03 & 9.41 (.001) & -0.03 \\
\hline
\multicolumn{5}{|c|}{Rydberg states}\\
\hline

$\pi$ $\rightarrow$ $\mathrm{3s}$ 
& 6.26 & -0.17 &  6.12 ($\mathrm{3s}$ /{\it w}+$\mathrm{3p_y}$)& -0.26 \\ 
& 6.28 &-0.16 &  6.17 ($\mathrm{3s}$ /{\it w}+ $\mathrm{3p_y}$)& -0.21 \\
$\pi_1$ $\rightarrow$ $\mathrm{3s}$ & 9.12 (.070) & -0.16&& \\
&&&& \\
&&&&\\
$\pi$ $\rightarrow$ $\mathrm{3p (p_x,p_y)}$ 

& 6.76 (.069) ($\mathrm{3p_y}$ /{\it w}) & -0.24 & 6.46 (.018) ($\mathrm{3p_y}$ /{\it w} +  $\mathrm{3s}$ /{\it w} +  $\mathrm{3d_{x^2-y^2}}$) & -0.38 \\
& 6.81             & -0.26    & {\it [6.56  ($\mathrm{3s}$ / w +  $\mathrm{3p_y}$)]}& -0.38 \\
& 6.82 (.005)         & -0.25  & 6.71 (.02) ($\mathrm{3p_x}$ +  $\mathrm{3d_{xy}}$) & -0.23\\
& 6.87 ($\mathrm{3p_y}$ /{\it w})     &  -0.27 & 6.77  ($\mathrm{3p_x}$ +  $\mathrm{3d_{xy}}$)& -0.28\\ 
$\pi$ $\rightarrow$ $\mathrm{3p (p_z)}$ 
& 6.89 (.119) & -0.32 & 6.82 (.21) ($\pi^\star$ +  $\mathrm{3d_{yz}}$)& -0.45 \\
& 6.92 (.147) & -0.30& 6.86 (.22) ($\pi^\star$ + $\mathrm{3d_{yz}}$)& -0.41\\
&   &&   & \\
 $\pi$ $\rightarrow$ $\mathrm{3d}$ 
& 7.29 (.002) (3d$_{z^2}$, $\mathrm{3d_{x^2-y^2}}$)& -0.37& 7.05 (.006) ($\mathrm{3d_{x^2-y^2}}$) &-0.47 \\
& 7.34 ($\mathrm{3d_{x^2-y^2}}$) &-0.33 & 7.11($\mathrm{3d_{x^2-y^2}}$)& -0.42 \\ 
& 7.41 ($\mathrm{3d_{xy}}$) &-0.32 & 7.23 ($\mathrm{3d_{xy}}$ )&-0.32\\
& 7.45 ($\mathrm{3d_{xy}}$) &-0.28 & 7.29 ($\mathrm{3d_{xy} + 3d_{z^2}}$ + $\mathrm{3p_x}$) &-0.26\\
&7.37 (d$_{z^2}$)&-0.20 & 7.25 ($\mathrm{3d_{z^2}}$)&-0.22\\
& 7.37 (d$_{z^2}$ + $\mathrm{3d_{x^2-y^2}}$) & -0.20 &7.28  ($\mathrm{3d_{z^2}}$) &-0.19\\
& 7.50  ($\mathrm{3d_{yz}}$)& -0.39 &7.31 ($\mathrm{3d_{yz}}$ + $\mathrm{3p_z}$+$\pi^\star$)& -0.39\\
& 7.55 ($\mathrm{3d_{yz}}$/$\mathrm{3d_{xz}}$) &-0.36& 7.37 (.007) ($\mathrm{3d_{xz}}$ + $\mathrm{3p_z}$ +  $\pi^\star$)&-0.44\\
 &7.57 (.002) ($\mathrm{3d_{xz}}$) &-0.34& 7.41 (.002) ($\mathrm{3d_{xz}}$)& -0.30\\
 &7.57 ($\mathrm{3d_{xz}}$/$\mathrm{3d_{yz}}$) &-0.37&7.43 ($\mathrm{3d_{xz}}$)& -0.32 \\
&&  && \\
 $\mathrm{n_O}$ $\rightarrow$ $\mathrm{(3s + 4s) /{\it w}}$
 &&& 7.20 (.22)&\\
 &&&& \\
\hline
 VIE & 8.91 && 8.91 (DFT)/9.07 (C-DFTB)&\\
 VIE2 & 8.94 &&& \\
 \hline
\end{tabular}
\caption{MS-CASPT2 and TD-DFT electronic excitation energies (in eV) for $Geo_{IED-1}$. The oscillator strengths are reported in parenthesis. The shifts of the transitions with respect to those in Bz are also reported ($\delta$E, in eV). The $\pi_1$ orbital refers to the $\pi (a_{2u})$ orbital in Bz. Vertical ionization energy (VIE) values are also reported.  
The transition in brackets and in italics refers to a "ghost state" as detected following the procedure indicated in Section \ref{subsec:compdct}. }
\label{tbl:Bz-H2O_geo3}
\end{table}

In summary, this study on the Bz-H$_2$O isomers shows the influence of the position and orientation of the water molecule on the spectrum of Bz. Although differences are observed between the MS-CASPT2 and the TD-DFT results due in particular to more valence Rydberg mixing and the description of transitions from $n_O$ orbitals in the latter case, the observed trends are similar. The positions of the split valence $\pi$ $\rightarrow$ $\pi^*$ intense transitions are very similar to that in isolated Bz in the case of $Geo_{IEI-1}$ while 
they are blue-shifted for $Geo_{IED-1}$ (7.10 instead of 6.94\,eV). Rydberg states are found at higher/lower energies for $Geo_{IEI-1}$/ $Geo_{IED-1}$. Of particular interest, low energy charge transfer states resulting from the excitation of one electron in a  $\pi$ orbital of  Bz to a Rydberg orbital located on the water molecule are observed for  $Geo_{IED-1}$ and not for $Geo_{IEI-1}$. \\

\subsection{Increasing the number of water molecules}\label{sub:Wn}

In this subsection, we consider the Bz molecule interacting with water clusters of increasing size following the two series of geometries $Geo_{IEI-n}$ and $Geo_{IED-n}$ as described in subsection\ref{subsec:comp1}. \\

\subsubsection{$Bz-(H_2O)_n$ : $Geo_{IEI-6}$ and $Geo_{IED-5}$}\label{subsec:BzW6-5}

\begin{figure}
        \centering
        \includegraphics[width=8cm]{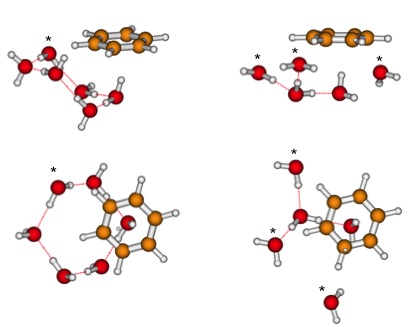} \\
        Geo$_{IEI-6}$ \hspace{4.5cm}Geo$_{IED-5}$ \\
        \caption{Geometries with the smallest shell of water molecules, extracted from the two Bz-Ih ice structures, one leading to the maximum ionization energy (Geo$_{IEI-n}$ series), the other one leading to the minimum ionization energy (Geo$_{IED-n}$ series).  The oxygen atoms pointing towards Bz are designated with the $\star$ symbol }
        \label{fig:geo_BzWm}
    \end{figure}

\begin{figure}
        \centering
       \includegraphics[width=14cm]{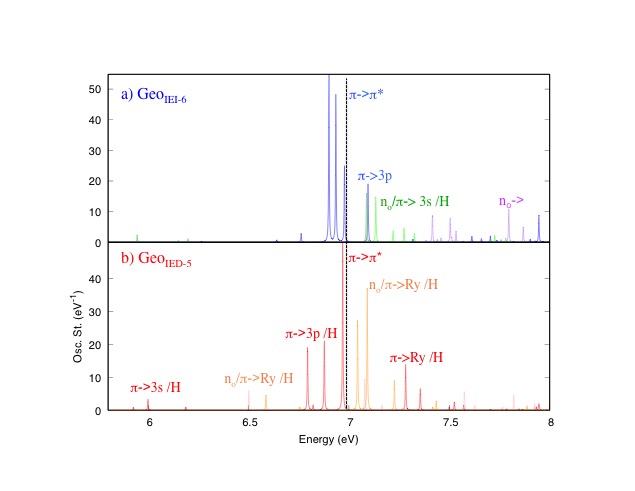} \\
        \vspace{-1.0cm}
         \includegraphics[width=14cm]{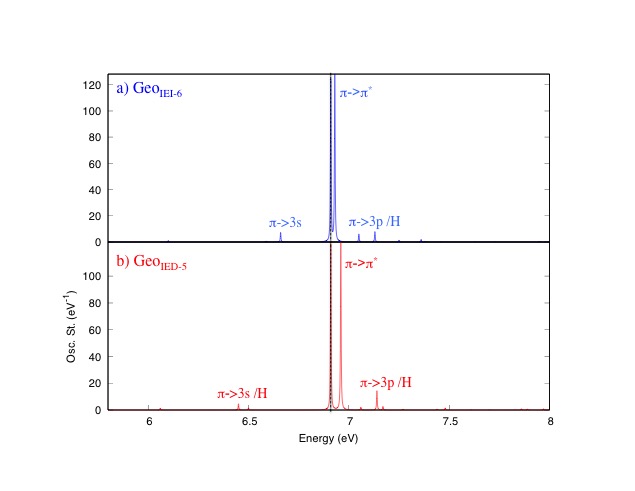} \\
        \caption{ TD-DFT (50 states, top)  and MS-CASPT2 spectra (bottom) 
        of   $Geo_{IEI-6}$ (a) and $Geo_{IED-5}$ (b). For TD-DFT spectra, the use of different colors allows to distinguish between transitions from $\pi$, $n_o$ and both ($n_o$, $\pi$) orbitals. The $\pi$-$\pi^{\star}$ transition energy for Bz is reported in dashed line.}
        \label{fig:spec2}
    \end{figure}

\begin{figure}
        \centering
        \includegraphics[width=7cm]{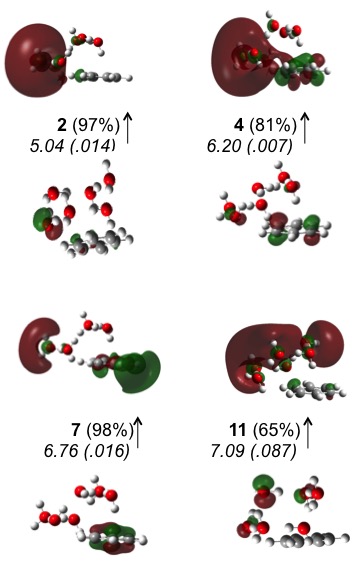}
        \caption{NTOs for lowest energy transitions involving electron transfer  to a Rydberg orbital located on the water cluster for Geo$_{IEI-6}$. The contours' isovalues are .08/.02 for the HOMOs/LUMOs respectively. Are also reported : the number of the transition, the contribution of the transition described by the NTOs with respect to the whole  excited state (in \%). The energy of the transition (in eV) and in oscillator strength are reminded in italics, the latter being in parenthesis.   }
        \label{fig:nto_34_6}
    \end{figure}

\begin{figure}
        \centering
        \includegraphics[width=7cm]{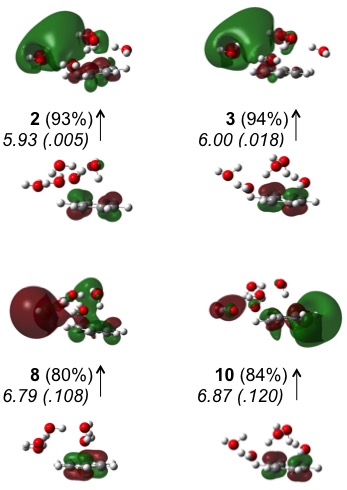}
       \caption{NTOs for lowest energy transitions involving electron transfer from the $\pi$ orbitals of benzene to a Rydberg orbital located on the water cluster for Geo$_{IED-5}$. The contours' isovalues are .08/.02 for the HOMOs/LUMOs respectively. Are also reported : the number of the transition, the contribution of the transition described by the NTOs with respect to the whole  excited state (in \%). The energy of the transition (in eV) and in oscillator strength are reminded in italics, the latter being in parenthesis. }
        \label{fig:nto_38_5}
    \end{figure}

In this section, we present the electronic excited states of Bz interacting with 5 and 6 water molecules ($Geo_{IED-5}$ and $Geo_{IEI-6}$) organized following the two configurations shown in Fig.~\ref{fig:geo_BzWm}. The MS-CASPT2 and TD-DFT electronic spectra are reported in Fig. \ref{fig:spec2}. The nature of the electronic transitions at both levels of theory is detailed in Tables \ref{tbl:Bz-H2O_34_6} and \ref{tbl:Bz-H2O_38_5}. Some NTOs of interest are  reported in  Figs.~\ref{fig:nto_34_6} and \ref{fig:nto_38_5}. The details of the assignments of the transitions, the MOs, as well as the D$_{\text{CT}}$ and  M$_{\text{AC}}$ indexes are reported in the SI (Geo\_IEI-6.pdf and  Geo\_IED-5.pdf files).\\

\newfloat{Figure}{H}{lof}

\newfloat{Figure}{H}{lof}

Interestingly, both $Geo_{IEI-6}$ and $Geo_{IED-5}$ were found to have similar VIEs at the C-DFTB level of theory. The VIEs of $Geo_{IEI-6}$ were found slightly smaller than those of $Geo_{IED-5}$ at the MS-CASPT2 level of theory (see last line of Tables~\ref{tbl:Bz-H2O_34_6} and \ref{tbl:Bz-H2O_38_5}). 
This is the opposite of the largest structures that they have been extracted from, likely because a water molecule, as in $Geo_{IEI-1}$, interacts with Bz through its H atom pointing  towards the $\pi$ cloud of Bz for the two structures.
As can be seen in Fig.~\ref{fig:spec2}, the TD-DFT electronic spectra present more transitions with non negligible oscillator strengths than at the MS-CASPT2 level. Indeed, several transitions from $n_O$ orbitals and from both $\pi$ and $n_O$ orbitals now come into play. For the two Bz-(H$_2$O)$_n$ clusters, the TD-DFT excitation energies of the water clusters without the Bz molecule were computed maintaining the presence of the Rydberg basis set, showing the presence of the expected bands (see Tables S2 and S3 of the SI). The difference in the excitation energies is larger for $Geo_{IED-5}$ with respect to $Geo_{IEI-6}$. The nature of the excitations has been carefully checked for the latter one and (H$_2$O)$_6$ while in $Geo_{IED-5}$ and (H$_2$O)$_5$ the mixing is too important to attribute with certainty the $n_O$ $\rightarrow$ $Ry$ transitions according to the contributions of the different oxygen atoms. However, from this comparison (SI, Tables S2 and S3), in both cases, the $n_O$ $\rightarrow$ $Ry$ transitions of the water clusters clearly appear in the Bz-clusters at similar energies (0.1-0.2\,eV).\\

Regarding the most intense $\pi \rightarrow \pi^{\star}$ transitions, their position is hardly affected but they undergo splitting due to symmetry loss. At the MS-CASPT2 and TD-DFT levels, this splitting leads to  the appearance of a band at slightly higher energy than that of Bz for $Geo_{IED-5}$ whereas for $Geo_{IEI-6}$, at the TD-DFT level, it leads to new bands at slightly lower energy. Regarding the values of the splitting, they were found identical (0.05\,eV) at the two levels of theory for $Geo_{IED-5}$ : the two bands are located at 6.91 and 6.96\,eV {\it vs} 6.91\,eV  for the isolated Bz at the MS-CASPT2 level, and at   6.97 and 7.04\,eV {\it vs} 6.99\,eV at the TD-DFT level. The results are slightly different for $Geo_{IEI-6}$ as hardly any splitting was found at the MS-CASPT2 level (6.91 and 6.93\,eV {\it vs} 6.91\,eV for Bz) and three bands were found at 6.90, 6.93 and 6.98\,eV {\it i. e.} -0.09, -0.06 and -0.01\,eV with respect to the band of isolated Bz. In fact, only two bands correspond to the $\pi \rightarrow \pi^{\star}$, but the three transitions are mixed. However, the band at 6.98\,eV presents a more pronounced $\pi \rightarrow Ry$ character than the two others.\\

We specify here that the analysis of the TD-DFT transitions was difficult. For these systems and all the more as the size of the water cluster increases, we found the lowest energy unoccupied MOs  developed on the Rydberg basis set also have significant contributions on diffuse orbitals and especially the 3s orbitals of H atoms (belonging to some water molecules).
This important contribution of the H atoms is quoted as /H in next tables and figures. The 4p atomic diffuse orbitals of the C and O orbitals, and to a lesser extent the 3s orbitals on the O atoms, also have non negligible contributions.  \\

Regarding $Geo_{IED-5}$, two $\pi$ $\rightarrow$ 3s /H transition at 6.45\,eV and 6.50\,eV were found at the MS-CASPT2 level of theory whereas those of similar nature were observed at the TD-DFT level at lower energies, 5.93\,and 6.00\,eV. In isolated Bz, these transitions were found at 6.33\,eV. This difference is one of the largest between the two methods. In $Geo_{IED-5}$, several transitions involve both $n_O$ and $\pi$ valence orbitals, as well as both Rydberg and $\pi^{\star}$ virtual orbitals. There are also two more intense $\pi$ $\rightarrow$ 3p /H at 6.79 and 6.87 eV, that we checked do not correspond to the population of ghost states (see SI, last table in the Geo\_IED-5.pdf file). Such transitions were observed at slightly higher energy at the MS-CASPT2 level (several bands around the most intense at 7.14\,eV), which is above the $\pi \rightarrow \pi^{\star}$ transition but remains much below the IP of Bz. In the case of $Geo_{IEI-6}$, $\pi \rightarrow Ry (3p)$ transitions were found slightly more intense than in the isolated benzene, at 6.98\,eV (mixed with $\pi \rightarrow \pi^{\star}$) and 7.10\,eV.  \\

\begin{figure}
        \centering
        \begin{tabular}{cc}
        \rotatebox{-90}{\includegraphics[width=5.0cm]{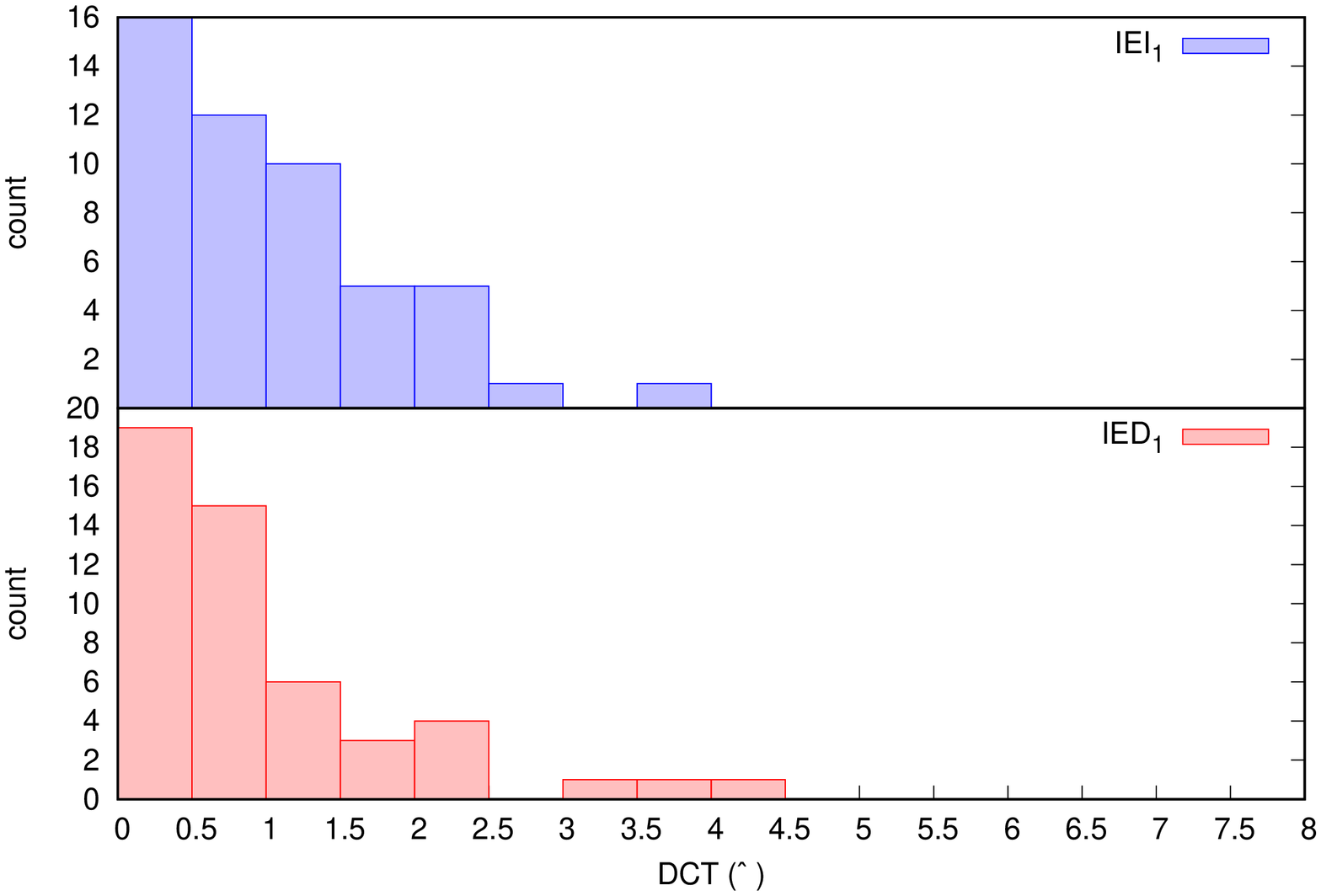}}& \rotatebox{-90}{\includegraphics[width=5.0cm]{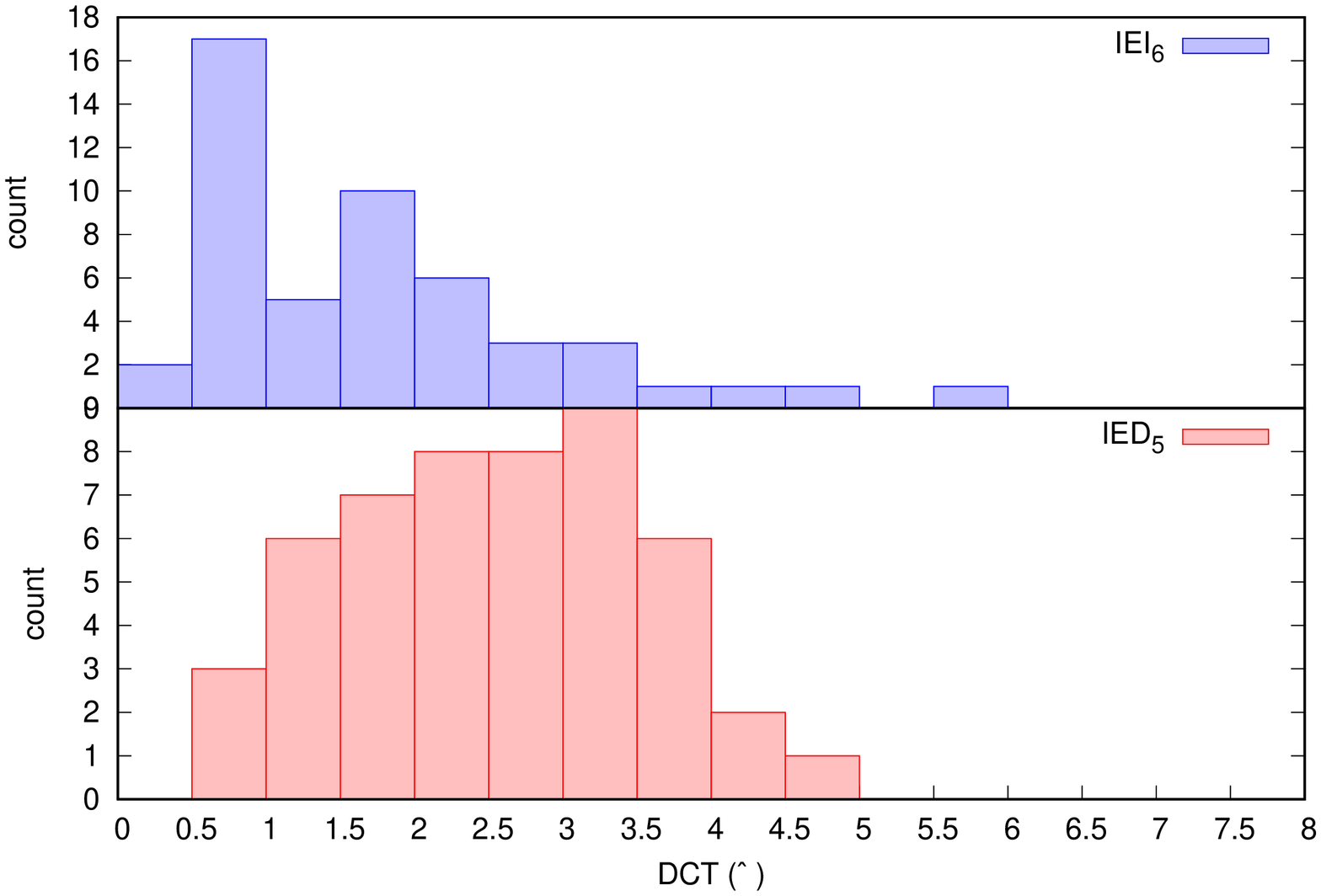}}\\
        \rotatebox{-90}{\includegraphics[width=5.0cm]{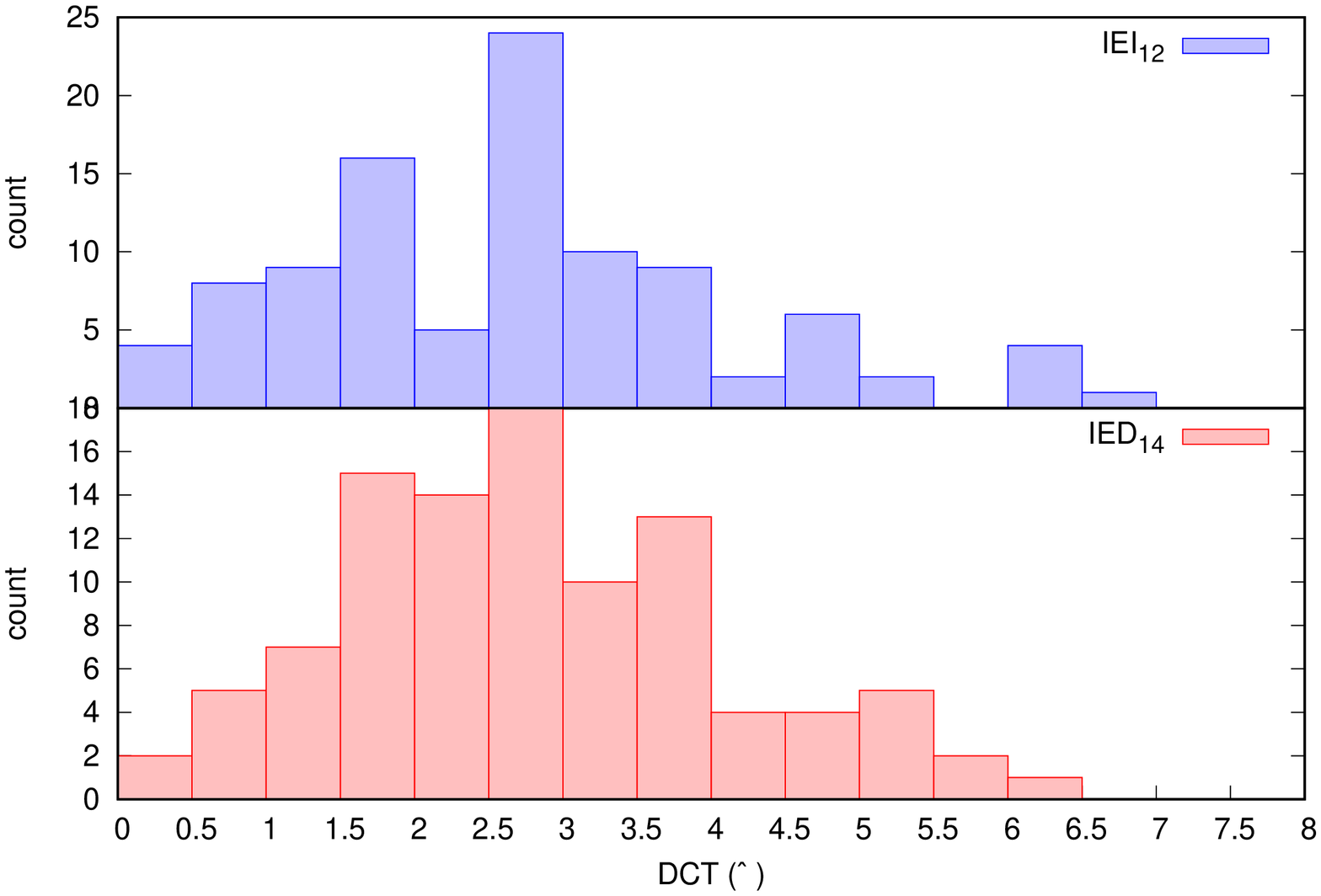}}& \rotatebox{-90}{\includegraphics[width=5.0cm]{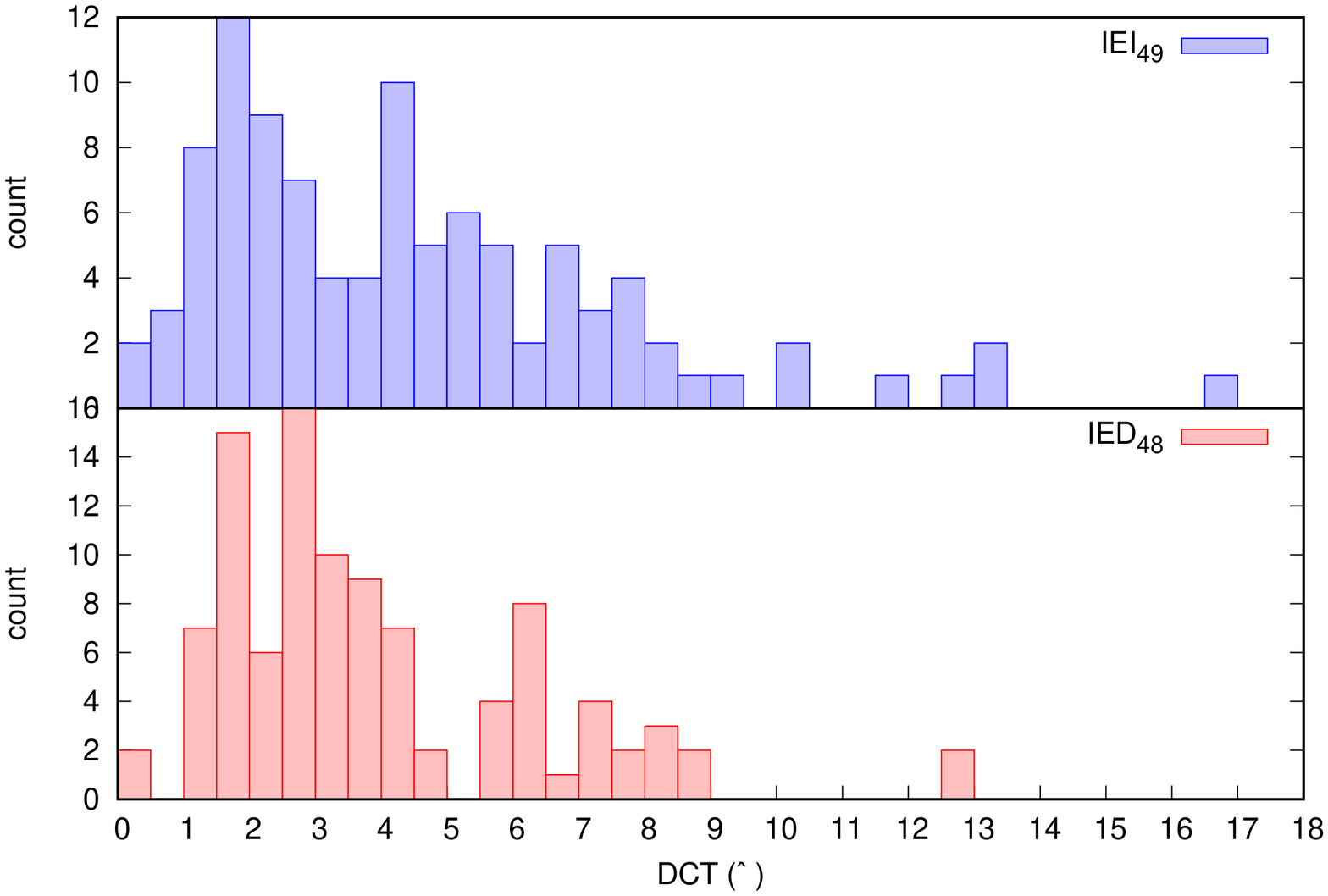}}\\
        \end{tabular}
       \caption{ D$_{\text{CT}}$ indexes' distributions for all clusters with more than one water molecule studied in this work (see the SI for the assignment of the D$_{\text{CT}}$ index to the corresponding transition).  }
       \label{fig:distrib_dct}
    \end{figure}

In regards with our goal of investigating charge transfer Bz$^+$-(H$_2$O)$_n$ states, the interesting result for such systems is the presence for $Geo_{IED-5}$, at both levels of theory, of $\pi \rightarrow Ry /H$  transitions of non negligible oscillator strength, at lower energy than the intense $\pi \rightarrow \pi^{\star}$ transition. As can be seen regarding the corresponding NTOs reported in Fig.~\ref{fig:nto_38_5}, that of transition 8 is developed in particular on the water molecule whose oxygen points towards an H atom of Bz, which can be regarded as the counterpart of transition 5 for $Geo_{IED-1}$ (see Fig.~\ref{fig:nto_geo3}). The D$_{\text{CT}}$ value for this transition is 3.22\,\AA\,, which falls into the maximum of the D$_{\text{CT}}$ values' distributions which are reported for all systems in Fig.~\ref{fig:distrib_dct}.  Interestingly, in the case of $Geo_{IEI-6}$, an analogous transition is transition 7 at 6.76\,eV (see Fig.~\ref{fig:nto_34_6}) but its D$_{\text{CT}}$ value is small (0.70\, \AA\, see SI, last table of Geo\_IEI-6.pdf file). 
Actually the D$_{\text{CT}}$ values for $Geo_{IEI-6}$ are globally smaller than the values of $Geo_{IED-5}$ (see Fig. \ref{fig:distrib_dct}). Besides, contrary to $Geo_{IED-5}$, some ghosts states were identified among the lowest energy transitions, and this is the case for the transition 2 reported in Fig.~\ref{fig:nto_34_6}
whereas that of transition 4 is not that clear (see SI, last table of Geo\_IEI-6.pdf file). \\


\newfloat{Table}{H}{lof}
\begin{table}
\begin{tabular}{|ll@{\hskip.8mm}|ll@{\hskip.8mm}|}
\hline
\multicolumn{4}{|c|}{$Geo_{IEI-6}$} \\
\multicolumn{2}{|c|}{MS-CASPT2/Gen. } &  \multicolumn{2}{c|}{TD-DFT/DunRy (2s2p2d1f) } \\
Transition & Energy & Transition  & Energy \\

\hline

$\pi$ $\rightarrow$ $\pi^\star$   &  4.98 (.001)  & $\pi$ $\rightarrow$ $\pi^\star$       &  5.47 (.001)  \\
                                 &               & $n_O$+$\pi$ $\rightarrow$ $Ry$ (3s)/ H  &  {\it [5.94 (.014)]}\\
$\pi$ $\rightarrow$ $\pi^\star$   &  6.10 (.006)  & $\pi$ $\rightarrow$ $\pi^\star$       &  6.15 (.003) \\
                                 &               & $\pi$ (+$n_O$) $\rightarrow$ $Ry$ (3s)/ H + $\pi$ $\rightarrow$ $\pi*$  &  6.20 (.007) \\
                                 &               & $\pi$ $\rightarrow$ $Ry$ (3s)/ H   &  {\it [ 6.26 (.002)]}\\
$\pi$ $\rightarrow$ $Ry$ (3s)    &  6.59 (.004)  & $\pi$ $\rightarrow$ $Ry$ (3p)   &  6.64 (.004)  \\
$\pi$ $\rightarrow$ $Ry$ (3s)    &  6.66 (.042)  & $\pi$ $\rightarrow$ $Ry$ (3p) + $Ry$ (3s)/ H   &  6.76 (.016) \\
$\pi$ $\rightarrow$ $\pi^\star$   &  6.91 (.723)  & $\pi$ $\rightarrow$ $\pi^\star$ + $Ry$ (3p) &   6.90 (.30)  \\
$\pi$ $\rightarrow$ $\pi^\star$   &  6.93 (.746)  & $\pi$ $\rightarrow$ $\pi^\star$  + $Ry$ (3p) &   6.93 (.26)\\
                                  &               & $\pi$ $\rightarrow$ $\pi^\star$  + $Ry$ (3p)   &  6.98 (.135) \\
$\pi$ $\rightarrow$ $Ry$ (3p$_y$) / H &  7.05 (.036)  & $n_O$ (+ $\pi$)  $\rightarrow$ $Ry$ (3s)/ H + $Ry$ (3p)  &  7.09 (.087) \\
 &    &  $\pi$ $\rightarrow$ $Ry$ (3p) + $\pi^\star$     &  7.10 (.11) \\
$\pi$ $\rightarrow$ $Ry$ (3p$_y$) / H &  7.13 (.046)  & $\pi$ $\rightarrow$ $Ry$ (3p) + $n_O$ $\rightarrow$ $Ry$ (3s)/ H  &  7.13 (.081)\\
$\pi$ $\rightarrow$ $Ry$ (3p$_x$)     &  7.25 (.008)  &   $\pi$ (+$n_O$)  $\rightarrow$ $Ry$ (s,p,d) + $\pi^\star$   &  7.22 (.022) \\
$\pi$ $\rightarrow$ $Ry$ (3p) / H &  7.36 (.011)  & $n_O$ (+ $\pi$ )$\rightarrow$ $Ry$ (3s)/ H + $Ry$ (3p)   &  7.27 (.025) \\
$\pi$ $\rightarrow$ $Ry$ (3p)  &  7.38 (.002)  & $\pi$ $\rightarrow$ $Ry$ (3p)  &  7.32 (.006)  \\
$\pi$ $\rightarrow$ $Ry$ (3p$_z$) / H &  7.46 (.001)  & $n_O$ + $\pi$ $\rightarrow$ $Ry$ (3s)/ H + $Ry$ (3p)  &  7.33 (.017) \\
$\pi$ $\rightarrow$ $Ry$ (3d$_{xy}$)    & 7.63 (.001)  & $n_O$ $\rightarrow$ $Ry$ (3p) + $\pi^\star$   &  7.38 (.004) \\
$\pi$ $\rightarrow$ $Ry$ (3d$_{xy}$)    & 7.73 (.001)  & $n_O$ $\rightarrow$ $Ry$ (3s)/ H  &  7.42 (.05) \\
$\pi$ $\rightarrow$ $Ry$ (3d$_{xz,yz}$)    & 7.81  & $\pi$ $\rightarrow$ $Ry$ (3d) +  $\pi^\star$  &  7.43 \\
$\pi$ $\rightarrow$ $Ry$ (3d$_{xz,yz}$)    & 7.91  & $n_O$ $\rightarrow$ $\pi^\star$ &  7.44 (.005) \\
$\pi$ $\rightarrow$ $Ry$ (3d$_{xz,yz}$)    & 7.95 (.002)  & $n_O$ $\rightarrow$ $\pi^\star$  &  7.46 (.008) \\
$\pi$ $\rightarrow$ $\pi^\star$ / $Ry$ (3d$_{z^2}$)   & 8.00 (.001)  & $n_O$ $\rightarrow$ $Ry$ (3s)/ H   &  7.50 (.042) \\
$\pi$ $\rightarrow$ $Ry$ (3d$_{xz,yz}$)    & 8.05 (.001)  & $n_O$ $\rightarrow$ $Ry$ (3s)/ H + $\pi^\star$  &  7.51 (.006) \\
$\pi$ $\rightarrow$ $Ry$ (3d$_{x^2-y^2}$)    & 8.08 (.001)  & $n_O$ $\rightarrow$ $Ry$ (3s)/ H + $Ry$ (3p) &  7.53 (.020) \\
$\pi$ $\rightarrow$ $Ry$ (3d$_{z^2}$)    & 8.10 (.001)  &                           &              \\
$\pi$ $\rightarrow$ $\pi^\star$  & 8.12 (.011)  &                           &              \\
$\pi$ $\rightarrow$ $Ry$ (3d$_{x^2-y^2}$)    & 8.17   &                           &              \\
$\pi$ $\rightarrow$ $\pi^\star$  & 8.19 (.002)  &                           &              \\
$\pi$ $\rightarrow$ $Ry$ (3s)   & 9.42 (.048)  &                           &              \\
\hline
VIE & 9.24 & VIE C-DFTB & 9.46\\
VIE2 & 9.34 &  & \\
 \hline
\end{tabular}
\caption{Analysis of the MS-CASPT2 and TD-DFT lowest energy transitions (in eV, up to the 25$^{th}$ excited state) for  
$Geo_{IEI-6}$. The oscillator strengths are reported in parenthesis. $Ry$ (3s)/ H refers to the LUMO orbital. The transitions in brackets and in italics refer to "ghost states" as detected following the procedure indicated in Section \ref{sec:comp}. }
\label{tbl:Bz-H2O_34_6}
\end{table}


\begin{table}
\begin{tabular}{|ll@{\hskip.8mm}|ll@{\hskip.8mm}|}
\hline
\multicolumn{4}{|c|}{$Geo_{IED-5}$} \\
\multicolumn{2}{|c|}{MS-CASPT2/Gen. } &  \multicolumn{2}{c|}{TD-DFT/DunRy (2s2p2d1f) } \\
Transition & Energy & Transition  & Energy \\
\hline
$\pi$ $\rightarrow$ $\pi^\star$   & 4.99              & $\pi$ $\rightarrow$ $\pi^\star$ [V+R]   & 5.47           \\
                                        &                   & $\pi$ $\rightarrow$ $Ry$ (3s) / H     & {\bf 5.93 (.005)}   \\
                                        &                   & $\pi$ $\rightarrow$ $Ry$ (3s) / H    & {\bf 6.00 (.018)}   \\
$\pi$ $\rightarrow$ $\pi^\star$   & 6.06 (.008)       & $\pi$ $\rightarrow$ $\pi^\star$  [V+R]  & 6.19 (.005)    \\
                                        &                   & $n_O$ $\rightarrow$ $Ry$  (3s + 3p) / H        & 6.50 (.033)  \\
                                        &                   & $\pi$  $\rightarrow$ $Ry$ (3p) / H + $n_O$  $\rightarrow$ $Ry$ (3s) / H   & 6.59 (.026)  \\
                                        &                   & $\pi$  $\rightarrow$ $Ry$ (3p) /H + $n_O$  $\rightarrow$ $Ry$ (3s) / H      & {\it [6.75 (.006)]} \\
$\pi$ $\rightarrow$ $Ry$ (3s) / H& {\bf 6.45 (.027)} & $\pi$ $\rightarrow$ $Ry$ (3p)    / H & {\bf 6.79 (.108)} \\
$\pi$ $\rightarrow$ $Ry$ (3s) / H& {\bf 6.50 (.009)} & $\pi$ $\rightarrow$ $Ry$ (3p + 3d)         & 6.82 (.009)  \\
                                        &                   & $\pi$ $\rightarrow$ $Ry$  (3p) / H    &  {\bf 6.87 (.120)} \\
$\pi$ $\rightarrow$ $\pi^\star$   & 6.91 (.727)       & $\pi$ $\rightarrow$ $\pi^\star$  [V+R]   & 6.97 (.284)   \\
$\pi$ $\rightarrow$ $\pi^\star$   & 6.96 (.736)       & $\pi$  $\rightarrow$ $\pi^\star$ [V+R] / $n_O$ $\rightarrow$ $Ry$ / H & 7.04 (.156)             \\
$\pi$ $\rightarrow$ $Ry$ (3p) / H   & {\bf 7.06 (.012)} & $n_O$ $\rightarrow$ $Ry$ / H       & 7.08 (.052)  \\
$\pi$ $\rightarrow$ $Ry$ (3p) / H   & {\bf 7.14 (.084)} & $\pi$ $\rightarrow$ $Ry$ [V+R] / $n_O$ $\rightarrow$ $Ry$ / H & 7.09 (.207)  \\
$\pi$ $\rightarrow$ $Ry$ (3p) / H   & {\bf 7.17 (.015)} & $n_O$ $\rightarrow$ $Ry$ / H        & 7.15 (.009)  \\
$\pi$ $\rightarrow$ $Ry$ (3p) / H   & {\bf 7.27 (.004)} & $n_O$ $\rightarrow$ $Ry$ /H + $\pi$  $\rightarrow$ $Ry$  (3p) / H  & 7.23 (.050) \\
$\pi$ $\rightarrow$ $Ry$ (3p)       & 7.44 (.004)       & $n_O$ $\rightarrow$ $Ry$  / H       & 7.26 (.003) \\
$\pi$ $\rightarrow$ $Ry$ (3p)       & 7.48 (.009)       & $\pi$ $\rightarrow$ $Ry$          & 7.28 (.075) \\
$\pi$ $\rightarrow$ $Ry$ (3d)       & 7.61 (.001)       & $\pi$ $\rightarrow$ $Ry$  / H + $\pi^\star$ [V+R]        & {\bf 7.35 (.036)} \\
$\pi$ $\rightarrow$ $Ry$ (3d)       & 7.70 (.001)       & $n_O$ $\rightarrow$ $Ry$  /  H (+ $\pi^\star$)  & 7.42 (.006) \\
$\pi$ $\rightarrow$ $Ry$ (3d)       & 7.83 (.001)       & $n_O$ $\rightarrow$ $Ry$  /  H (+ $\pi^\star$)   & 7.44 (.016) \\
$\pi$ $\rightarrow$ $Ry$ (3d)       & 7.86 (.005)       & $\pi$ $\rightarrow$ $Ry$ /  H + $\pi^\star$  [V+R] & {\bf 7.50 (.006)} \\
$\pi$ $\rightarrow$ $Ry$ (3d)       & 7.89 (.004)       & $\pi$ $\rightarrow$ $Ry$ /  H + $\pi^\star$  [V+R] & 7.53 (.014)  \\
$\pi$ $\rightarrow$ $Ry$ (3d)       & 7.97 (.005)       & $n_O$ $\rightarrow$ $Ry$ /  H + $\pi^\star$ [V+R] & 7.54  \\
$\pi$ $\rightarrow$ $Ry$ (3d)       & 8.00              & $\pi$ $\rightarrow$ $Ry$ /  H + $\pi^\star$ [V+R] & {\bf 7.57 (.008)} \\
$\pi$ $\rightarrow$ $Ry$ (3d)       & 8.04 (.001)       &  & \\
$\pi$ $\rightarrow$ $Ry$ (3d)       & 8.06              &                                   &                \\
$\pi$ $\rightarrow$ $Ry$ (3d)       & 8.10              &                                   &                \\
$\pi$ $\rightarrow$ $\pi^\star$     & 8.17 (.014)       &                                   &                \\
$\pi$ $\rightarrow$ $\pi^\star$     & 8.19 (.001)       &                                   &                \\
$\pi$ $\rightarrow$ $Ry$ (3s) / H   & 9.23 (.050)       &                                   &               \\

    \hline
VIE  ($\pi$)                        & 9.32              & VIE                               & 9.45 (C-DFTB)\\
VIE2 ($\pi$)                        & 9.38              &                                   &                \\
 \hline
\end{tabular}
\caption{Analysis of the MS-CASPT2 and TD-DFT lowest energy transitions (in eV, up to the 25$^{th}$ excited state) for $Geo_{IED-5}$. The oscillator strengths are reported in parenthesis. The Ry(3s/3p) orbitals refer to the LUMO/LUMO+1 orbitals. 
The important contribution of the diffuse orbitals located on the H atoms of the water molecules is indicated by " / H". When mixtures between valence and Rydberg contributions are significant, [V+R] is added. The transitions in brackets and in italics refer to "ghost states" as detected following the procedure indicated in Section \ref{subsec:compdct}. }
\label{tbl:Bz-H2O_38_5}
\end{table}

    \subsubsection{$Bz-(H_2O)_n$ : $Geo_{IEI-12}$ and $Geo_{IED-14}$}

\begin{figure}
        \centering
        \includegraphics[width=10cm]{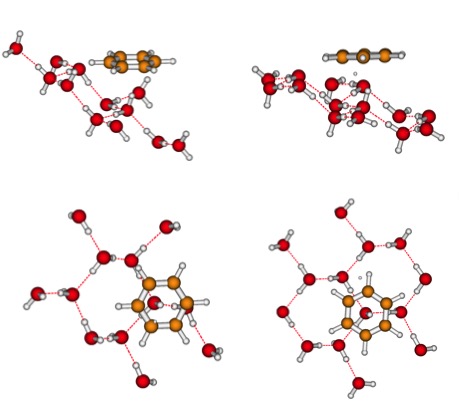} \\
         Geo$_{IEI-12}$ \hspace{4.5cm}Geo$_{IED-14}$ \\
        \caption{Geometries extracted from the Geo$_{IEI-n}$ (left) and Geo$_{IED-n}$ (right) series containing respectively 12 and 14 water molecules. } 
        \label{fig:geo_BzWl}
    \end{figure}

\begin{figure}
       \centering
       \includegraphics[width=14cm]{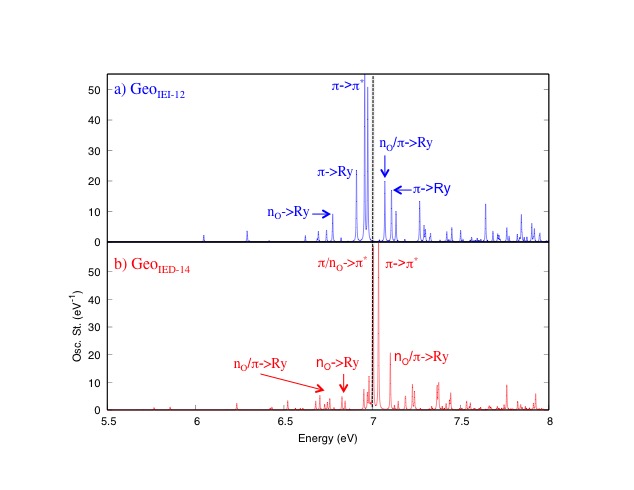}
        \vspace{-1cm}
        \caption{TD-DFT electronic spectra (100 states) for $Geo_{IED-14}$   (red) and  $Geo_{IEI-12}$  (blue). }
        \label{fig:sp_large}
    \end{figure}

\begin{figure}
        \centering
        \includegraphics[width=7cm]{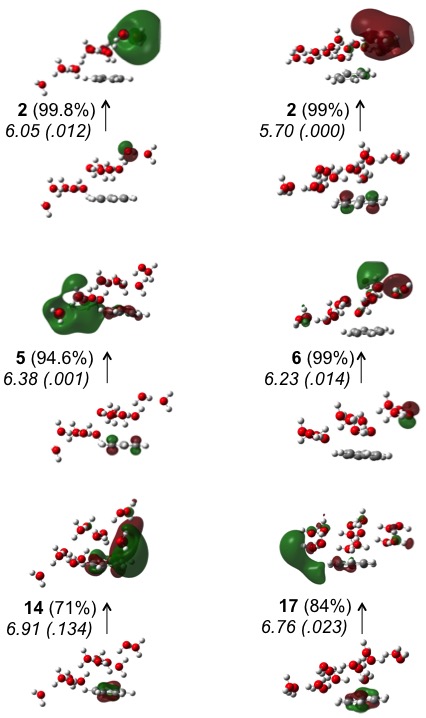} \\
        $Geo_{IEI-12}$ \hspace{3.5cm} $Geo_{IED-14}$ \\
        \caption{NTOs related to a few low energy transitions for $Geo_{IEI-12}$ (left) and $Geo_{IED-14}$ (right). The HOMOs/LUMOS are drawn with a contour of .08/.02 respectively. Are also reported : the number of the transition, the contribution of the transition described by the NTOs with respect to the complete excited state (in \%). The energy of the transition (in eV) and in oscillator strength are reminded in italics, the latter being in parenthesis.    }
        \label{fig:nto_12_14}
    \end{figure}

Increasing the size of the water clusters leads to  $Geo_{IEI-12}$ and  $Geo_{IED-14}$ shown in Fig.~\ref{fig:geo_BzWl}. The TD-DFT electronic spectra are reported in Fig.~\ref{fig:sp_large}, and the assignments of the lowest energy transitions for both clusters are reported in Table \ref{tbl:Bz-H2O-12-14} based on the transitions' detailed assignments and MOs detailed in the SI (Geo\_IEI-12.pdf and Geo\_IED-14.pdf files).  A few NTOs are shown in Fig.~\ref{fig:nto_12_14}. \\

Increasing the number of water molecules as specified in subsection \ref{subsec:comp1} has hardly any effect on the $\pi \rightarrow \pi^{\star}$ transitions which are split but whose positions remain close to that of Bz (less than .05\,eV deviation). They are slightly blue-shifted in the case of $Geo_{IED-14}$, one of them resulting from both $\pi$ and $n_O$ orbitals' transition to  a $\pi^{\star}$ orbital, and slightly red-shifted in the case of  $Geo_{IEI-12}$. These bands are surrounded by more low intensity bands than in $Geo_{IED-5}$ and $Geo_{IEI-6}$ due to the increased number of transitions from  $n_O$ orbitals and  from both $\pi$ and $n_O$  orbitals. \\

For both $Geo_{IED-14}$ and $Geo_{IEI-12}$, low energy transitions can be regarded as contributing to the population of Bz$^+$- (H$_2$O)$_n^-$ charge transfer states, although many of them are of negligible oscillator strength. However, for both structures, a few low lying excited states involving $\pi \rightarrow Ry /H$ transition of non negligible oscillator strength are involved, as for instance transition 14 for $Geo_{IEI-12}$ and transition 17 for $Geo_{IED-14}$, whose NTOs are reported in Fig.~\ref{fig:nto_12_14},  and whose energies remain below 7.0\,eV, that is to say more than 2\,eV below the IP of Bz.  Interestingly, the MOs referred to as Rydberg orbitals also have contribution on the diffuse H atoms of the water molecules located at the edge of the water cluster, leading to globally larger D$_{\text{CT}}$ values. 
As for the other systems, the possibility of the presence of ghost states was investigated and the only suspected ghost state would be the 3$^{rd}$ one for $Geo_{IED-14}$ (see the SI, Geo\_IEI-14.pdf file).  The D$_{\text{CT}}$ value for this state is among the largest one for $Geo_{IED-14}$ (5.37 \AA), the D$_{\text{CT}}$ values for transitions 6 and 17 were found at $\sim$ 3 \AA, those for transitions 2, 5 and 14 of $Geo_{IEI-12}$ at 2.45, 3.41 and 1.13 \AA\, respectively (for unrelaxed densities). Overall, as can be seen in Fig.~\ref{fig:distrib_dct}, the D$_{\text{CT}}$ values for $Geo_{IED-14}$ and $Geo_{IEI-12}$ are more spreaded towards larger distances than those of $Geo_{IED-5}$ and $Geo_{IEI-6}$, illustrating the longer distances browsed by the electron upon excitation. \\

 {\tiny
 \begin{table}
 \begin{tabular}{|l|l@{\hskip.8mm}|l|l@{\hskip.8mm}|}
\hline
\multicolumn{2}{|c|}{$Geo_{IED-14}$} & \multicolumn{2}{|c|}{$Geo_{IEI-12}$} \\
Transition                       & Energy  & Transition                        & Energy  \\
\hline
$\pi$ $\rightarrow$ $\pi^\star$         & 5.49          & $\pi$ $\rightarrow$ $\pi^\star$     & 5.48 \\
$\pi$ $\rightarrow$ $Ry 3s$             & 5.70          &                                     &       \\
$n_O$ / $\pi$ $\rightarrow$ $Ry$        & 5.76          &                                     &       \\
$n_O$ / $\pi$ $\rightarrow$ $Ry$        & 5.86 (.01)    & $n_O$ (HOMO) $\rightarrow$ $Ry$     & 6.05 (.01)   \\
$\pi$ $\rightarrow$ $\pi^\star$         & 6.17          & $\pi$ $\rightarrow$ $\pi^\star$     & 6.17 \\
$n_O$ $\rightarrow$ $Ry$                & 6.23 (.01)    & $n_O$ $\rightarrow$ $Ry$            & 6.29 (.02) \\
$n_O$ $\rightarrow$ $Ry$                & 6.42          & $\pi$ $\rightarrow$ $Ry$            & 6.31  \\
$\pi$ $\rightarrow$ $Ry$                & 6.43          & $\pi$ $\rightarrow$ $Ry$            & 6.42  \\
$n_O$ $\rightarrow$ $Ry$                & 6.52 (.02)    & $n_O$ / $\pi$ $\rightarrow$ $Ry$    & 6.62 (.01) \\
$n_O$ / $\pi$ $\rightarrow$ $Ry$        & 6.57          & $n_O$ (HOMO) $\rightarrow$ $Ry$     & 6.69   \\
$n_O$ / $\pi$ $\rightarrow$ $Ry$        & {\it [6.59]}  & $n_O$ / $\pi$ $\rightarrow$ $Ry$    & 6.70 (.02) \\
$n_O$ / $\pi$ $\rightarrow$ $Ry$        & 6.61          &                                     &       \\
$n_O$ / $\pi$ $\rightarrow$ $Ry$        & 6.68 (.02)    &                                     &       \\
$n_O$ / $\pi$ $\rightarrow$ $Ry$        & 6.70 (.03)    &                                     &       \\
$n_O$ / $\pi$ $\rightarrow$ $Ry$        & 6.73 (.01)    &                                     &       \\
$n_O$ / $\pi$ $\rightarrow$ $Ry$        & 6.75 (.02)    &                                     &       \\
$\pi$ / $n_O$ $\rightarrow$ $Ry$        & 6.76 (.02)    & $\pi$ $\rightarrow$ $Ry$            & 6.74  \\
$n_O$ / $\pi$ $\rightarrow$ $Ry$        & 6.78          & $n_O$ $\rightarrow$ $Ry$            & 6.78 (.05)  \\
$n_O$ $\rightarrow$ $Ry$                & 6.83 (.03)    & $n_O$ $\rightarrow$ $Ry$            & 6.82 (.01) \\
$n_O$ $\rightarrow$ $Ry$                & 6.85 (.02)    & $n_O$ (HOMO) $\rightarrow$ $Ry$     & 6.86  \\     
$\pi$ $\rightarrow$ $Ry$                & 6.87          & $\pi$ $\rightarrow$ $Ry$            & 6.91 (.13)  \\
$n_O$ $\rightarrow$ $Ry$                & 6.95 (.04)    &                                     &       \\
$n_O$ $\rightarrow$ $Ry$                & 6.97 (.03)    &                                     &       \\
$n_O$ / $\pi$ $\rightarrow$ $Ry$        & 6.98 (.06)    &                                     &       \\
$n_O$ / $\pi$ $\rightarrow$ $Ry$        & 6.99 (.01)    &                                     &       \\
$\pi$ / $n_O$ $\rightarrow$ $\pi^\star$ & 7.00 (.32)    & $\pi$ $\rightarrow$ $\pi^\star$     & 6.96 (.30) \\
$\pi$ $\rightarrow$ $\pi^\star$         & 7.04 (.33)    & $\pi$ $\rightarrow$ $\pi^\star$     & 6.97 (.27) \\
                                        &               & $n_O$ (HOMO) $\rightarrow$ $Ry$     & 7.04  \\
$n_O$ / $\pi$ $\rightarrow$ $Ry$        & 7.10 (.11)    & $n_O$ / $\pi$ $\rightarrow$ $Ry$    & 7.07 (.11) \\
$\pi$ $\rightarrow$ $Ry$                & 7.13 (.01)    & $\pi$ $\rightarrow$ $Ry$            & 7.11 (.10) \\
$n_O$ $\rightarrow$ $Ry$                & 7.15 (.02)    & $\pi$ $\rightarrow$ $Ry$            & 7.13 (.06) \\
$n_O$ $\rightarrow$ $Ry$                & 7.19 (.02)    &                                     &       \\
$n_O$ $\rightarrow$ $Ry$                & 7.19 (.02)    &                                     &       \\
$n_O$ $\rightarrow$ $Ry$                & 7.23 (.05)    &                                     &       \\
$\pi$ $\rightarrow$ $Ry$                & 7.24 (.04)    & $\pi$ $\rightarrow$ $Ry$            & 7.18 \\
$n_O$ $\rightarrow$ $Ry$                & 7.25          & $n_O$ (HOMO) $\rightarrow$  $\pi^\star$ & 7.27 (.01) \\
$n_O$ $\rightarrow$ $Ry$ / $\pi^\star$  & 7.27          &                                     &       \\
$n_O$ / $\pi$ $\rightarrow$ $Ry$        & 7.32          & $\pi$ $\rightarrow$ $Ry$            & 7.27 (.06) \\
$n_O$ $\rightarrow$ $Ry$                & 7.34          & $n_O$ $\rightarrow$ $Ry$            & 7.29 (.03) \\
$n_O$ $\rightarrow$ $Ry$                & 7.34          & $n_O$ (HOMO) $\rightarrow$ $Ry$     & 7.30 (.02) \\
$n_O$ $\rightarrow$ $Ry$                & 7.36          & $n_O$ (HOMO) $\rightarrow$ $\pi^\star$ & 7.33 (.01) \\
$n_O$ $\rightarrow$ $Ry$                & {\it [7.37 (.03)]}    & $\pi$ $\rightarrow$ $Ry$    & 7.33 (.02) \\
$n_O$ $\rightarrow$ $Ry$                & 7.38 (.04)    & $n_O$ $\rightarrow$ $Ry$            & 7.38  \\
$n_O$ $\rightarrow$ $Ry$                & 7.40          & $n_O$ (HOMO) $\rightarrow$ $\pi^\star$ & 7.41       \\
$\pi$ / $n_O$ $\rightarrow$ $Ry$        & {\it [7.41]}          & $n_O$ $\rightarrow$ $Ry$    & 7.42 (.02) \\
                                        &               & $n_O$ / $\pi$ $\rightarrow$ $Ry$    & 7.43 \\
                                        &               & $n_O$ / $\pi$ $\rightarrow$ $Ry$    & 7.45 (.03) \\
                                        &               & $n_O$ / $\pi$ $\rightarrow$ $Ry$ and $\pi^\star$& 7.50(0.02) \\
                                        &               & $n_O$ $\rightarrow$ $Ry$            & 7.51       \\
                                        &               & $\pi$ $\rightarrow$ $Ry$            & 7.55 \\
                                        &               & $n_O$ / $\pi$ $\rightarrow$ $Ry$ and $\pi^\star$& 7.56(0.01) \\
                                        &               & $\sigma_{CH}$ $\rightarrow$ $\pi^\star$ & 7.58 \\
                                        &               & $n_O$ $\rightarrow$ $Ry$            & 7.58 \\
                                        &               & $n_O$ $\rightarrow$ $Ry$            & 7.59 (.01) \\
                                        &               & $n_O$ $\rightarrow$  $\pi^\star$    & 7.60       \\
                                        &               & $n_O$ $\rightarrow$ $Ry $           & 7.62 \\
                                        &               & $n_O$ $\rightarrow$ $Ry$            & 7.64 (.07) \\
                                        &               & $\pi$ $\rightarrow$ $Ry$            & 7.68 (0.02) \\ 

\hline
VIE                                     & 8.93 (C-DFTB) & VIE                                 & 9.57 (C-DFTB)\\
\hline
 \end{tabular}
\caption{Analysis of the TD-DFT/DunRy (2s2p2d1f) lowest energy transitions (in eV) for  $Geo_{IED-14}$ up to the 40$^{th}$ excited state) and  $Geo_{IEI-12}$ (up to the 43$^{th}$ excited state). The transitions in brackets and in italics refer to "ghost states" as detected following the procedure indicated in Section \ref{subsec:compdct}. The oscillator strengths are reported in parenthesis. } 
\label{tbl:Bz-H2O-12-14}
\end{table}
 }

This study of intermediate-size systems confirms the dependence of the electronic spectrum on the organisation of the water molecules interacting with Bz. Interestingly, in the view of identifying low energy Bz$^+$-[(H$_2$O)$_n$]$^-$ CT states, our calculations show that  a few real $\pi$ $\rightarrow$ Ry /H states with non negligible oscillator strengths are present for Geo$_{IED-14}$ and Geo$_{IED-12}$.\\

    \subsubsection{$Bz-(H_2O)_n$ : $Geo_{IEI-49}$ and $Geo_{IED-48}$}\label{subsec:vl}

\begin{figure}
        \centering
        \includegraphics[width=14cm]{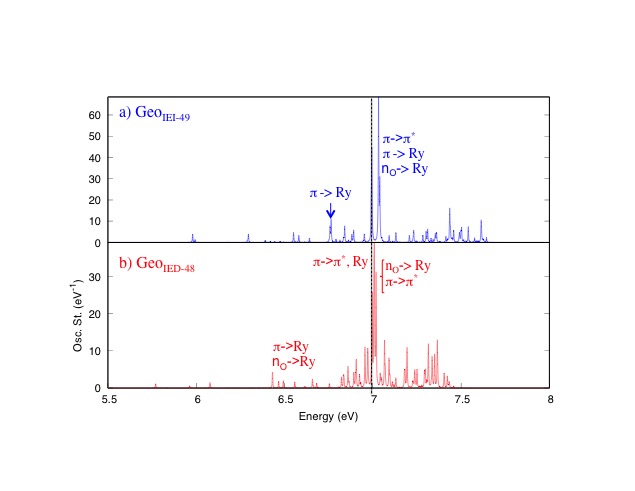}
       \vspace{-2cm}
        \caption{TD-DFT electronic spectra (100 states)  of $Geo_{IED-48}$   (bottom red) and  $Geo_{IEI-49}$  (top blue). }
        \label{fig:sp_big}
    \end{figure}

\begin{figure}
        \centering
        \includegraphics[width=10cm]{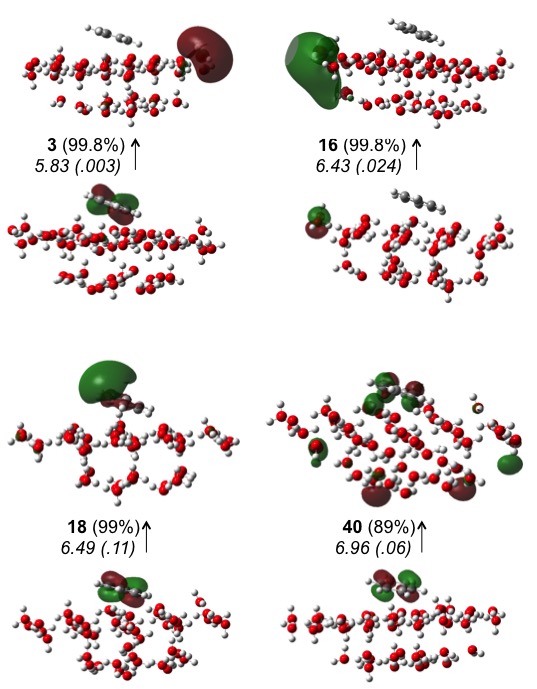}
        \caption{Examples of NTOs involved in transitions from $\pi$  and $n_O$ orbitals to Rydberg orbitals for $Geo_{IED-48}$. The contours' isovalues are .05/0.02 for HOMOS and LUMOS respectively. The number of the transition, the contribution of the transition described by the NTOs with respect to the complete excited state (in \%) are also reported. The energy of the transition (in eV) and oscillator strength are reminded in italics, the latter being in parenthesis.}
        \label{fig:nto_38-48}
    \end{figure}

\begin{figure}
        \centering
        \includegraphics[width=10cm]{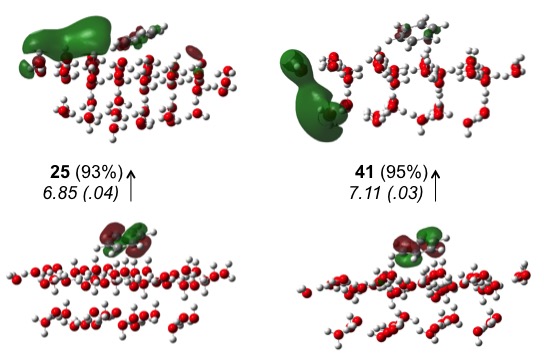}
        \caption{Examples of NTOs involved in transitions from $\pi$ to Rydberg orbitals for $Geo_{IEI-49}$. The contours' isovalues are .05/.02 for the HOMOs/LUMOs respectively. The number of the transition, the contribution of the transition described by the NTOs with respect to the complete excited state (in \%) are also reported. The energy of the transition (in eV) and oscillator strength are reminded in italics, the latter being in parenthesis. A more complete set for the NTOs of $Geo_{IEI-49}$ are reported in the SI (Geo\_IEI\_49.pdf file). }
        \label{fig:nto_34-49}
    \end{figure}

    The electronic spectra of $Geo_{IEI-49}$ and $Geo_{IED-48}$ 
are shown in Fig.~\ref{fig:sp_big} and the assignment of the lowest energy transitions are reported in Table \ref{tbl:Bz-H2O-48-49}. A few NTOs describing $\pi$ $\rightarrow$ Ry /H transitions are pictured in Figs. \ref{fig:nto_38-48} and  \ref{fig:nto_34-49}. All transitions' detailed assignments, MOs and D$_{\text{CT}}$ and M$_{\text{AC}}$ indexes are detailed in the SI (see files Geo\_IEI-49.pdf and Geo\_IED-48.pdf).\\
    
    On the electronic spectra, we observe intense $\pi \rightarrow \pi^{\star}$ dominant transitions slightly blue-shifted - by less than .05 eV - with respect to that of Bz. Hardly any splitting (less than .02 eV) was found for Geo$_{IEI-48}$, 0.04\,eV for Geo$_{IEI-49}$. These intense excitations are not pure $\pi \rightarrow \pi^{\star}$ as the latter are combined with $\pi \rightarrow Ry $ and $n_O \rightarrow Ry $ transitions. Interestingly, regarding these transitions, our results differ from those obtained by Sharma et al. \cite{Sharma2016_bzIce} who computed a red-shift of these bands with respect to those within isolated Bz when increasing the size of the water clusters, up to describing the Ih water ice environment using a QM/MM appproach. However, as previously mentioned in the introduction, they used a different functional from ours (M06-2X, that is not a long range corrected functional) and a basis set that does not describe the Rydberg states of Bz.    
    These intense bands are surrounded by "weeds" {\it ie} numerous bands of weak intensities with oscillator strengths about ten times weaker, describing the promotion of electrons from $n_O$ and $\pi$ orbitals to Rydberg orbitals. Among those, some of them are suspected to be ghost states (see SI). This is the case for instance of low energy $\pi \rightarrow Ry$ transitions ranging from 5.84 to 6.05\,eV for Geo$_{IED-48}$. These suspected ghost states are of negligible oscillator strength. \\
   As previously mentionned in the case of smaller clusters, the MOs having a contribution on Rydberg atomic orbitals also have contribution on diffuse orbitals, in particular on the H atoms of water molecules located at the edge of the water cluster (see for instance the LUMO of the transitions reported in Fig.~\ref{fig:nto_38-48}). This is illustrated by the spreading of the D$_{\text{CT}}$ index towards long distances (see Fig.~\ref{fig:distrib_dct}) up to 17 \AA. A representative case is for instance the charge transfer $\pi \rightarrow Ry/H$ transition 3 of Geo$_{IED-48}$ which is characterized by a D$_{\text{CT}}$ value of 7.1 \AA. Transitions 36 and 40, more intense, are characterized by D$_{\text{CT}}$ values of 2.6 and 3.8 \AA.
   For the two geometries, low energy $\pi \rightarrow Ry$ transitions are present and can be regarded as charge transfer Bz$^+$-[(H$_2$O)$_n$]$^-$ states. As mentioned previously, and contrary to the smallest clusters (n=5,6) the Rydberg states for such large systems are mostly located on the edge of the water clusters and are developed on both the Bz Rydberg orbitals and on the atomic diffuse orbitals located mostly on the H atoms of the water molecules at the edge of the cluster.  \\

 {\tiny
\begin{table}

 \begin{tabular}{|l|l@{\hskip.8mm}|l|l@{\hskip.8mm}|}
\hline
\multicolumn{2}{|c|}{$Geo_{IED-48}$} & \multicolumn{2}{|c|}{$Geo_{IEI-49}$} \\
Transition                       & Energy  & Transition                        & Energy  \\
\hline
                                                                  &             & $n_O$ / H $\rightarrow$ $Ry$                                    &  5.28         \\
$\pi$ $\rightarrow$ $\pi^\star$                                   & 5.50        & $\pi$ $\rightarrow$ $\pi^\star$                                    &  5.49         \\
$n_O$ (HOMO-2) $\rightarrow$ $Ry$                                    & 5.77 (.01)  & $n_O$ $\rightarrow$ $Ry$                                           &  5.99 (0.02)  \\
$\pi$ $\rightarrow$ $Ry$                                          & 5.84        & $n_O$ $\rightarrow$ $Ry$                                           &  6.02 (0.01)  \\
$\pi$ $\rightarrow$ $Ry$                                          & 5.90        & $\pi$ $\rightarrow$ $\pi^\star$                                    &  6.18         \\
$\pi$ $\rightarrow$ $Ry$                                          & 5.93        & $n_O$ /H $\rightarrow$ $Ry$                                       &  6.33         \\
$n_O$ (HOMO-2) $\rightarrow$ $Ry$                                    & 5.96        & $n_O$ $\rightarrow$ $Ry$                                           &  6.33 (0.02)  \\
$\pi$ $\rightarrow$ $Ry$                                          & 5.97        & $n_O$ $\rightarrow$ $Ry$                                           &  6.41 (0.01)  \\
$\pi$ $\rightarrow$ $Ry$                                          & 6.02        & $n_O$ $\rightarrow$ $Ry$                                           &  6.42         \\
$\pi$ $\rightarrow$ $Ry$                                          & 6.05        & $n_O$ /H $\rightarrow$ $Ry$                                       &  6.45         \\
$n_O$ $\rightarrow$ $Ry$                                          & 6.08 (.01)  & $\pi$ $\rightarrow$ $Ry$                                           &  6.47         \\
$\pi$ $\rightarrow$ $\pi^\star$                                   & 6.18        & $n_O$ /H $\rightarrow$ $Ry$                                       &  6.51         \\
$\pi$ $\rightarrow$ $Ry$                                          & 6.23        & $n_O$ $\rightarrow$ $Ry$                                           &  6.52 (0.02)  \\
$\pi$ $\rightarrow$ $Ry$                                          & 6.28        & $\pi$ $\rightarrow$ $Ry$                                           &  6.55         \\
$\pi$ $\rightarrow$ $Ry$                                          & 6.32        & $n_O$ /H $\rightarrow$ $Ry$                                       &  6.62 (0.02)  \\
$\pi$ $\rightarrow$ $Ry$                                          & 6.36        & $n_O$ $\rightarrow$ $Ry$                                           &  6.63 (0.01)  \\
$n_O$ $\rightarrow$ $Ry$                                          & 6.43 (.02)  & $n_O$ $\rightarrow$ $Ry$                                           &  6.64         \\
$\pi$ $\rightarrow$ $Ry$                                          & 6.47 (.01)  & $n_O$ /H $\rightarrow$ $Ry$                                       &  6.67 (0.01)  \\
$\pi$ $\rightarrow$ $Ry$                                          & 6.49 (.01)  & $n_O$ $\rightarrow$ $Ry$                                           &  6.71 (0.03)  \\
$n_O$ $\rightarrow$ $Ry$                                          & 6.56 (.01)  & $\pi$ $\rightarrow$ $Ry$                                           &  6.77 (0.08)  \\
$n_O$ $\rightarrow$ $Ry$                                          & 6.58        & $n_O$ $\rightarrow$ $Ry$                                           &  6.80 (0.01)  \\
$\pi$ $\rightarrow$ $Ry$                                          & 6.62        & $n_O$ $\rightarrow$ $Ry$                                           &  6.81         \\
$n_O$ $\rightarrow$ $Ry$                                          & 6.66 (.01)  & $n_O$ /H $\rightarrow$ $Ry$                                       &  6.83         \\
$\pi$ $\rightarrow$ $Ry$                                          & 6.66        & $n_O$ $\rightarrow$ $Ry$                                           &  6.84         \\
$n_O$ $\rightarrow$ $Ry$                                          & 6.68 (.01)  & $\pi$ $\rightarrow$ $Ry$                                           &  6.85 (0.04)  \\
$\pi$ $\rightarrow$ $Ry$                                          & 6.75 (.01)  & $n_O$ $\rightarrow$ $Ry$                                           &  6.86 (0.01)  \\
$\pi$ $\rightarrow$ $Ry$                                          & 6.81        & $n_O$ $\rightarrow$ $Ry$                                           &  6.87 (0.02)  \\
$n_O$ $\rightarrow$ $Ry$                                          & 6.82 (.02)  & $n_O$ $\rightarrow$ $Ry$                                           &  6.89 (0.03)  \\
$n_O$ $\rightarrow$ $Ry$ / $\pi$ $\rightarrow$ $Ry$               & 6.83        & $n_O$ /H $\rightarrow$ $Ry$                                       &  6.91         \\
$n_O$ $\rightarrow$ $Ry$ / $\pi$ $\rightarrow$ $Ry$               & 6.83 (.01)  & $n_O$ $\rightarrow$ $Ry$                                           &  6.94         \\
$n_O$ $\rightarrow$ $Ry$                                          & 6.84 (.01)  & $n_O$ $\rightarrow$ $Ry$                                           &  6.97 (0.04)  \\
$n_O$ $\rightarrow$ $Ry$                                          & 6.86 (.03)  &   $\pi$ $\rightarrow$ $\pi^\star$                                 &  7.00 (0.31)  \\
$\pi$ $\rightarrow$ $Ry$                                          & 6.87 (.01)  & $n_O$ $\rightarrow$ $Ry$                                    &  7.00 (0.01)  \\
$n_O$ (HOMO-2) $\rightarrow$ $Ry$                                    & 6.88        & $n_O$ $\rightarrow$ $Ry$                                           &  7.02         \\
$n_O$ $\rightarrow$ $Ry$                                          & 6.89 (.02)  & $n_O$ /H $\rightarrow$ $Ry$ / $\pi^\star$                         &  7.03         \\
$n_O$ $\rightarrow$ $Ry$                                          & 6.90 (.02)  & $\pi$ $\rightarrow$ $\pi^\star$ / $\pi$ $\rightarrow$ $Ry$                      &  7.04 (0.37)  \\
$\pi$ $\rightarrow$ $Ry$                                          & 6.91 (.04)  & $n_O$ $\rightarrow$ $Ry$ / $\pi$ $\rightarrow$ $\pi^\star$         &  7.04 (0.11)  \\
$n_O$ (HOMO-2) $\rightarrow$ $Ry$                                    & 6.92        & $n_O$ /H $\rightarrow$ $Ry$ / $n_O$ /H $\rightarrow$ $\pi^\star$ &  7.08         \\
$\pi$ $\rightarrow$ $Ry$                                          & 6.92 (.02)  & $n_O$ $\rightarrow$ $Ry$                                           &  7.10 (0.02)  \\
$n_O$ $\rightarrow$ $Ry$                                          & 6.93 (.01)  & $n_O$ /H $\rightarrow$ $Ry$ / $n_O$ /H $\rightarrow$ $\pi^\star$ &  7.10         \\
$\pi$ $\rightarrow$ $Ry$                                          & 6.96 (.06)  & $\pi$ $\rightarrow$ $Ry$                                           &  7.11 (0.02)  \\
$n_O$ $\rightarrow$ $Ry$                                          & 6.97 (.04)  & $n_O$ /H $\rightarrow$ $\pi^\star$                                &  7.13         \\
$n_O$ $\rightarrow$ $Ry$                                          & 6.97 (.03)  & $n_O$ $\rightarrow$ $Ry$                                           &  7.15         \\
$\pi$ $\rightarrow$ $Ry$                                          & 6.98        & $n_O$ $\rightarrow$ $Ry$                                           &  7.17         \\
$\pi$ $\rightarrow$ $Ry$ / $\pi$ $\rightarrow$ $\pi^\star$        & 7.00 (.14)  & $n_O$ $\rightarrow$ $Ry$                                           &  7.18         \\
$n_O$ $\rightarrow$ $Ry$                                          & 7.00        & $\pi$ $\rightarrow$ $Ry$                                           &  7.20 (0.03)  \\
$n_O$ (HOMO-2) $\rightarrow$ $Ry$ / $\pi$ $\rightarrow$ $\pi^\star$  & 7.01 (.12)  & $n_O$ $\rightarrow$ $Ry$                                           &  7.21 (0.03)  \\
$n_O$ (HOMO-2) $\rightarrow$ $Ry$ / $\pi$ $\rightarrow$ $\pi^\star$  & 7.01 (.16)  &                                                                    &           \\
$\pi$ $\rightarrow$ $Ry$ / $\pi$ $\rightarrow$ $\pi^\star$        & 7.02 (.16)  &                                                                    &           \\

\hline
VIE                              & 8.40 (C-DFTB) & VIE                                        & 9.80 (C-DFTB)\\
\hline
 \end{tabular}
\caption{Analysis of the TD-DFT/DunRy (2s2p2d1f) lowest energy transitions (in eV)  for  $Geo_{IED-48}$ (up to the 48$^{th}$ excited state) and  $Geo_{IEI-49}$ (up to the 46$^{th}$ excited state). The oscillator strengths are reported in parenthesis.} 
\label{tbl:Bz-H2O-48-49}
\end{table}
 }

   \section{Discussion}\label{sec:disc}
   Our results show the presence of low energy (less than 7.0\,eV)  charge transfer  Bz$^+$-(H$_2$O)$_n^-$ states with non negligible oscillator strength for the smallest clusters when the oxygen of a water molecule points towards an H atom of Bz. The orbitals receiving the electron is an orbital which has a Rydberg character but which is also developed on atomic diffuse orbitals, in particular those of the H atoms of some water molecules. Their nature is then different from the $\pi \sigma ^{\star}$ CT states evoked by Noble {\it et al.} to account for the photoreactivity of coronene:water clusters \cite{Noble2017} as no antibonding contribution has been observed in the populated orbitals of the excited states. The correlation between the relative orientation of the water molecule and the existence of a Bz$^+$-(H$_2$O)$_n^-$ CT state, that was clearly identified when the  oxygen of a water molecule interacts with an H atom of Bz, is consistent with the previous results based on combined experimental and theoretical studies \cite{simon_formation_2017}. Indeed, these studies show that, in an argon matrix, the most stable structures of a coronene molecule in interaction with a water monomer or dimer correspond to structures with one water molecule interacting through its oxygen atom with an H atom of the PAH in the PAH plane. Therefore such structures, that are not the most stable in the gas phase, would be the reactive species in the argon matrix experiments, leading to the photoactivated oxydation of the PAHs up to the formation of quinones \cite{Guennoun2011,Guennoun2011bis}. \\ 
    
    When the size of the cluster increases, the Rydberg orbitals expand and also contain diffuse orbitals located on the H atoms of water molecules at the edge of the water clusters.  Longer range transitions are involved, which can be seen on the D$_{\text{CT}}$ distributions. Interestingly, not many ghost states were clearly found. Their nature is often ambiguous as the energy of the excited states is in the vast majority of the cases less than 1.5\,eV below the M$_{\text{AC}}$ index. We may note that the significance of such a criterion was shown for large delocalized $\pi$ systems \cite{campetella_charge_2017} and it is the first time that it is used for such molecular clusters so the conclusions concerning the ghosts states must be regarded with caution. \\

    Overall, CT states of non negligible oscillator strength can be found below 7\,eV {\it ie} less than 2\,eV below the ionisation potential of isolated Bz when the latter interacts with water clusters. When these water clusters are small, our results such states are present when a water monomer interacts with the H atoms of Bz via its O atom. Interestingly, such conformations also tend to lead to a lowering of the IE of Bz.  When the size of the cluster increases, it seems that such states are present for any configurations, including those leading to an increase of the IE of Bz. 
    However, we might wonder about the lifetime of such states and about their reactivity. In this respect, radical processes were also invoked for the reactivity of PAH with water \cite{campetella_charge_2017}, which goes beyond the study presented in this work.\\
    
    Finally, we may wonder whether the results obtained in the present paper for finite size systems can be applied to Bz adsorbed on or embedded in water ice. Considering large water clusters, we showed that the electron transferred to the  cluster within the CT states was preferentially located on the H atoms at the edge of the cluster. If we extrapolate our results to the Bz adsorbed on/embedded in water ice, that would mean that the transferred electron is preferentially located on the H atoms pertaining to dOH bonds that could be present either in some cavities where Bz could be trapped in or on the ice surface where Bz could be adsorbed on. Interestingly, recent combined experimental and theoretical studies based on the dOH infrared diagnosis showed that the location of Bz in ice was not clear and that it depended on the Bz:water relative concentration \cite{michoulier_perturbation_2020}.  Interestingly, the fate of the electron determined with our approach appears quite in line with the current models, however still debated, used to describe the solvated electron. Indeed, it seems that the majority of such models  are consistent with a cavity picture in which the excess electron density is self-trapped, localized and primarily contained in a solvent-free void within the water \cite{turi_theoretical_2012}, which could correspond to the dOH at the surface or inside some cavities (defaults) of interstellar water ice.

    \section{Conclusion}\label{sec:conc}
  
    In this work, we presented 
    the electronic spectra and the detailed analysis of the transitions up to $\sim$ 8\,eV 
    of a Bz molecule in interaction with water clusters of different structural organisations and different sizes.  These were extracted from two benzene-Ih ice configurations leading to maximum/minimum ionization energies 
    of the solvated benzene. 
    For all systems, the electronic spectra were computed at the TD-DFT level in conjunction with an appropriate basis set containing diffuse and polarisation orbitals on the atoms and describing the Rydberg states of benzene. The approach was carefully benchmarked against MS-CASPT2 results for the smallest systems.
    
     Transitions from the non bonding $n_O$ orbitals are described at the TD-DFT level, which is not the case of MS-CASPT2 spectra due to the restricted active space. TD-DFT transitions were found to be extremely multireferential, and a transition might involve both $n_O$ and $\pi$ orbitals as well as both and Rydberg and $\pi^{\star}$ virtual orbitals. Despite some discrepancies, the trends were found to be similar with both MS-CASPT2 and TD-DFT methods: the positions and intensities of the main $\pi \rightarrow \pi^{\star}$ transitions were determined to be hardly affected and only splitting due to symmetry breaking was observed, whose value is similar at the two levels of theory. 
     When the size of the cluster increased, an increasing number of weak  transitions at both lower and higher energy than these intense transitions were found, a large number of these being transitions from $n_O$ and $n_O$/$\pi$ orbitals to Rydberg orbitals. \\
     
    For the smallest systems, our results clearly demonstrate the influence of the clusters' structures on the spectra. In particular, only for the $Geo_{IED}$ series, we found a $\pi \rightarrow $Ry /H transition of non negligible intensity describing the promotion of one electron to a Rydberg orbital having an important contribution on the water molecule whose oxygen points towards the H atom of Bz. For larger systems, a $\pi \rightarrow $Ry /H transition below 7\,eV was observed for all configurations. The difference with the small systems is that the Rydberg orbitals become mainly developed on the H atoms of the water molecules at the edge of the  cluster. 
    Such $\pi \rightarrow $Ry /H transitions, that can be regarded as CT states, were found more than 2\,eV below the IP of Bz. As a result, such states could be formed irradiating the systems under low energy photons as achieved in the photoreactivity experiments \cite{Guennoun2011,Guennoun2011bis}. However, in order to rationalize the experimental results {\it ie} the formation of a long lived CT state and its reactivity, further investigations such as non adiabatic dynamics are necessary. \\
    
    \section{Supplementary Information (SI)}\label{sec:SI}
   A folder named {\bf SI\_Bz\_water\_TCA} contains 9 files described hereafter : 
    \begin{itemize}
        \item {\bf SI\_bench\_Bz\_water.pdf }contains all the SI refering to the benchmmark of the electronic spectrum  the benzene molecule and of the water clusters as well as Rydberg basis sets generated for MS-CASPT2 calculations.
        \item the 8 other files named {\bf Geo\_IEI-1.pdf}, {\bf Geo\_IED-1.pdf}, {\bf Geo\_IEI-6.pdf}, {\bf Geo\_IED-5.pdf}, {\bf Geo\_IEI-12.pdf},  {\bf Geo\_IED-14.pdf}, {\bf Geo\_IEI-49.pdf} and  {\bf Geo\_IED-48.pdf} contain the geometry (in xyz), the basis set, the TD-DFT energies and excitations' detailed assignments, the shape of the MOs, a list of the transitions with the corresponding D$_{\text{CT}}$ and M$_{\text{AC}}$ indexes. The few of the two latter values which were computed using the relaxed densities are also reported. In the case of {\bf Geo\_IEI-49.pdf} and  {\bf Geo\_IEI-12.pdf}, several NTOs are also shown.
    \end{itemize}
    
    \section{Acknowledgments}
    The authors would like to thank J. Mascetti, C. Toubin and J. Noble for fruitful scientific discussions. The authors are also deeply grateful to F. Spiegelman for his dedication to Science and his tremendous scientific expertise he has always been eager to share with them. A.S. also acknowledges the computing facility CALMIP for generous allocation of computing resources (project P17002) at University of Toulouse (UT 3). This work has been funded by the French Agence Nationale de la Recherche (ANR) project “PARCS” ANR-13-BS08-0005, with support of the French research network EMIE (Edifices Mol\'eculaires Isol\'es et Environn\'es, GDR 3533), and of the French National Program “Physique et Chimie du Milieu Interstellaire” (PCMI) of the CNRS/INSU with the INC/INP, co-funded by the CEA and the CNES.

\bibliographystyle{spphys}

\end{document}